\documentclass[letterpaper,journal]{IEEEtran}
\usepackage{amsmath,amsfonts}
\usepackage{algorithmic}
\usepackage{algorithm}
\usepackage{array}
\usepackage[caption=false,font=normalsize,labelfont=sf,textfont=sf]{subfig}
\usepackage{textcomp}
\usepackage{stfloats}
\usepackage{url}
\usepackage{verbatim}
\usepackage{graphicx}
\usepackage{cite}
\usepackage{enumitem} 

\usepackage{xcolor}
\usepackage{amssymb}
\usepackage[colorlinks=true, linkcolor=black, citecolor=blue, urlcolor=blue]{hyperref} 
\usepackage[capitalise,nameinlink]{cleveref}
\crefformat{figure}{\textcolor{blue}{Fig.\,#2\textcolor{blue}{#1}#3}} 
\crefmultiformat{figure}
  {\textcolor{blue}{Figs.\,}#2\textcolor{blue}{#1}#3}
  {\textcolor{blue}{~and~}#2\textcolor{blue}{#1}#3}
  {\textcolor{blue}{,~}#2\textcolor{blue}{#1}#3}
  {\textcolor{blue}{, and~}#2\textcolor{blue}{#1}#3}
\Crefname{figure}{\textcolor{blue}{Fig.}}{\textcolor{blue}{Figs.}} 
\crefformat{equation}{\textcolor{blue}{Eq.}~#2\textcolor{blue}{(#1)}#3} 
\crefmultiformat{equation}
  {\textcolor{blue}{Eqs.~}#2\textcolor{blue}{(#1)}#3}
  {\textcolor{blue}{~and~}#2\textcolor{blue}{(#1)}#3}
  {\textcolor{blue}{,~}#2\textcolor{blue}{(#1)}#3}
  {\textcolor{blue}{, and~}#2\textcolor{blue}{(#1)}#3}
\Crefname{equation}{\textcolor{blue}{Eq.}}{\textcolor{blue}{Eqs.}} 
\crefformat{table}{\textcolor{blue}{Table~#2\textcolor{blue}{#1}#3}}
\crefmultiformat{table}
  {\textcolor{blue}{Tables~}#2\textcolor{blue}{#1}#3}
  {\textcolor{blue}{~and~}#2\textcolor{blue}{#1}#3}
  {\textcolor{blue}{,~}#2\textcolor{blue}{#1}#3}
  {\textcolor{blue}{, and~}#2\textcolor{blue}{#1}#3}
\Crefname{table}{\textcolor{blue}{Table}}{\textcolor{blue}{Tables}}
\usepackage{doi}        
\usepackage{booktabs}   
\usepackage{siunitx}    
\DeclareSIUnit{\sample}{Sa}

\begin{document}
\title{Terahertz Radar Inversion for Range-Resolved Solids Concentration Profiling in Gas--Solid Flows}
\author{P.~K.~Jepsen, A.~Monteith, D.~C.~Guío-Pérez, L.~Ulander, T.~Bryllert, H.~Ström, and D.~Pallarès%
\thanks{P.~K.~Jepsen, D.~C.~Guío-Pérez, and D.~Pallarès are with the Division of Energy Technology, Department of Environmental and Energy Sciences, Chalmers University of Technology, SE-412 96 Gothenburg, Sweden.}%
\thanks{H.~Ström is with the Division of Fluid Dynamics, Department of Mechanical Engineering, Chalmers University of Technology, SE-412 96 Gothenburg, Sweden.}%
\thanks{T.~Bryllert is with the Division of Terahertz and Millimetre Wave Laboratory, Department of Microtechnology and Nanoscience, Chalmers University of Technology, SE-412 96 Gothenburg, Sweden.}%
\thanks{L.~Ulander and A.~Monteith are with the Division of Geoscience and Remote Sensing, Department of Environmental and Energy Sciences, Chalmers University of Technology, SE-412 96 Gothenburg, Sweden.}%
\thanks{Corresponding author: P. K. Jepsen (e-mail: jepsen@chalmers.se).}%
}

\markboth{Preprint submitted to IEEE Transactions on Instrumentation and Measurement}{}

\IEEEaftertitletext{%
\vspace{0.5\baselineskip}
\begin{center}
\footnotesize\itshape
This work has been submitted to the IEEE for possible publication. Copyright may be transferred without notice, after which this version may no longer be accessible.
\end{center}
\vspace{0.5\baselineskip}
}

\maketitle

\begin{abstract}
A stable inversion method is developed to reconstruct millimeter-scale solids-concentration profiles in particulate suspensions from monostatic sub-terahertz frequency-modulated continuous-wave (FMCW) radar measurements. In industrial particulate flows, cumulative attenuation renders the direct inversion of the path-integrated volume-scattering radar equation ill-conditioned. The method resolves this by an ensemble-averaged Mie-scattering closure coupled to a stable backward-integration scheme, yielding an analytical solution for the range-resolved solids concentration without case-specific parameter fitting against reference concentration measurements. The required range-dependent system response is determined by absolute near- and far-field calibration via external substitution with an electrically large metallic sphere. The method was evaluated against solids-volume-fraction estimates derived from differential-pressure measurements in a 3.1 m-tall circulating fluidized-bed riser operated with ambient air and copper powder (median diameter 32.6\;$\boldsymbol{\mu}$\text{m},\; density 8920\;\text{kg\,m$\boldsymbol{^{-3}}$}). The radar-derived and pressure-derived profiles agreed well across three superficial gas velocities producing distinct axial solids distributions, and the radar indicates sensitivity down to solids volume fractions of order 10$\boldsymbol{^{-6}}$.
\end{abstract}


\section{Introduction}
\IEEEPARstart{A}{ccurate} measurements of solids flow and concentration in gas--solid systems, such as fluidized beds, spouted beds, and pneumatic conveying lines, are required for process optimization and for establishing quantitative descriptions of solids-transport mechanisms \cite{FBreactorProcessIntensification2024}. These flows exhibit heterogeneous structures, including particle clustering, void formation, and rapid fluctuations in solids distribution across multiple spatial and temporal scales \cite{Gomez_Leckner_BiomassModellingReview_2010,ToveHenrik_DNS}. Predictive modeling of such multiphase systems, for example within Eulerian--Eulerian or Eulerian--Lagrangian frameworks \cite{XIONG2013_Eulerian,CFDmodellingDenseSolidsFlow2025}, requires constitutive closure relations for unresolved interphase momentum exchange, particle--particle interactions, and mesoscale heterogeneities. Because such closures are typically empirical, spatially and temporally resolved measurements of particle distributions, mixing patterns, and solids mass fluxes are needed to constrain model assumptions, support reactor design, and identify undesirable operating regimes \cite{Zhu2024GasSolidsReview,NonInvasiveMethods_Review_2024,ExpAndSimCharacterizationCFBs2026}.

Many diagnostic techniques have been used to quantify solids holdup (volume fraction) in gas--solid systems, and several reviews are available in the literature \cite{NonInvasiveMethods_Review_2024,Zhu2024GasSolidsReview,ReviewPaper_ECT_in_Pharma_WANG2021,ReviewPaper_ECT_in_CFBs_WANG2020}. Intrusive local probes (e.g.\ optical-fiber, capacitance, and electrostatic probes) are simple and low cost, but provide point measurements and can perturb the flow \cite{ReviewMeasurementFluidizedBeds2008, Werther_2000_CapacitanceProbes, LI202064_electrostatic_probes}. Differential-pressure measurements between vertically spaced wall taps provide cross-sectionally averaged solids-holdup estimates, but sensitivity decreases at low holdup and taps may clog \cite{SASICBoLecknerFilipJohnsson_2007,ReviewMeasurementFluidizedBeds2008}. Tomography techniques can reconstruct cross-sectional solids distributions. Electrical capacitance tomography (ECT), widely used in fluidized beds, is safe to operate and provides frame rates of tens to hundreds of frames per second, but its spatial resolution is typically limited to 5\%--10\% of the cross-sectional diameter and degrades toward the center. This limitation arises from the non-uniform electric-field sensitivity of the circumferential electrode array and the resulting nonlinear ill-posed inverse problem of reconstructing the cross-sectional solids distribution \cite{ReviewPaper_ECT_in_Pharma_WANG2021,ReviewMeasurementFluidizedBeds2008,ReviewIndustrialMixMonitoringTechniques2020,NonInvasiveMethods_Review_2024}. X-ray and $\gamma$-ray tomography offer higher spatial resolution, but typically at lower frame rates, while radiation safety requirements and facility constraints limit deployment \cite{ReviewPaper_ECT_in_Pharma_WANG2021,ReviewPaper_ECT_in_CFBs_WANG2020,XrayReview2025}.

In radar remote sensing, electromagnetic waveforms are transmitted, and the received scattered field is processed to obtain range-resolved profiles along the beam \cite{Dey2022ParticleSpectroscopyReview}. In the past decade, high-frequency radar remote sensing of distributed electromagnetic scatterers has been applied to diagnostics of solids flow in particulate multiphase systems of industrial relevance \cite{RadarForSolids_2016_Reinhardt,RadarForSolids_2017_Reinhardt,RadarBased_Tomas_2023_IEEE,ThzRadar_AstraZeneca_2024,Radar_Solids_System_2025,THzRadarOtherGroupBioMass2025}.

The need for finer spatial resolution has motivated interest in terahertz (\SIrange{0.3}{3}{\tera\hertz}) radars, whose large bandwidths and short wavelengths support millimeter-scale range resolution and increased sensitivity to submillimeter particles through narrow beams and high antenna gain for a given aperture \cite{SubMillimeterWaveRadar_DesignAndApplication_2014,RadarBased_Tomas_2023_IEEE}. The main constraints are the limited availability and high cost of THz components and the strong atmospheric attenuation above \SI{0.3}{\tera\hertz}, which increases rapidly with frequency and restricts the practical measurement range \cite{SubMillimeterWaveRadar_DesignAndApplication_2014,subTHzTradeOffsCommercially_2024}.

For short-range sensing (up to \SI{6}{\meter}), a compact monostatic frequency-modulated continuous-wave (FMCW) pulse--Doppler radar with a center frequency of \SI{0.34}{\tera\hertz} and a fractional bandwidth of 9.4\% was developed \cite{RadarBased_Tomas_2023_IEEE} and demonstrated in dilute particle streams \cite{RadarBased_Marlene_2023_AdvancedPowderTechnology,ThzRadar_AstraZeneca_2024} and in a circulating fluidized bed (CFB) containing glass beads with a mean size of \SI{106}{\micro\meter} \cite{RadarBased_Carolina_2023_CFBs,RadarBased_Wu_2024}. Range--Doppler maps were obtained with \SI{5}{\milli\meter} range resolution and \SI{0.03}{\meter\per\second} velocity resolution, from which local particle-velocity statistics can be estimated. 

Quantitative retrieval of solids concentration from the backscattered power requires inversion of a forward electromagnetic model relating solids concentration to the measured radar return. The forward model in \cite{RadarBased_Marlene_2023_AdvancedPowderTechnology,RadarBased_Carolina_2023_CFBs,RadarBased_Wu_2024} uses a single-wavelength path-integrated radar equation under a single-scattering assumption to relate radar return power to solids concentration. Under the strong cumulative attenuation encountered in gas--solid flow systems, direct inversion becomes severely ill-conditioned. Previous work addressed this by fitting two parameters in the backscatter and extinction terms to obtain stable solutions matched to solids volume fractions estimated from pressure measurements. This prevents independent validation of the inversion and restricts application to configurations where concurrent reference concentration measurements are available, which limits transfer across operating conditions, measurement positions, and viewing angles. The model basis also remains uncertain because potential near-field effects have not been addressed, and the approximately 10\% fractional bandwidth requires justification of the single-wavelength narrowband approximation. 

The resulting question is how strongly--attenuated terahertz radar backscatter can support quantitative range-resolved solids-concentration retrieval without case-specific parameter fitting against reference concentration measurements. The aim of this work is to develop an analytical inversion procedure that remains stable under strong cumulative attenuation, together with the range-dependent calibration, beam characterization, and ensemble-averaged scattering closure required for the retrieval.

To address this question, this work employs a methodology with the following contributions:
\begin{enumerate}[label=\Roman*.]
    \item An analytical inversion of the path-integrated radar equation is derived and evaluated using stabilized backward integration from a far-range boundary to reconstruct range-resolved solids-concentration profiles from strongly attenuated terahertz radar backscatter. The required far-range boundary value is estimated directly from the measured radar return under an attenuation-dominated local approximation.

    \item A range-dependent absolute calibration procedure based on an electrically large metallic reference sphere is established to characterize the system response, including antenna beam characteristics, over the relevant measurement interval.

    \item An ensemble-averaged scattering closure is formulated from the measured polydisperse particle-size distribution and Mie theory. For the present waveform bandwidth, the intraband frequency dependence of the scattering coefficients is small relative to the variation induced by particle-size polydispersity, supporting a narrowband approximation.
\end{enumerate}
The inversion framework is evaluated for vertical measurements in a \SI{3.1}{\meter}-tall CFB riser containing copper powder fluidized by ambient air. \footnote{The copper powder was selected to satisfy the fluid-dynamic scaling criteria of the cold-flow facility relative to an industrial \SI{330}{\mega\watt_{\mathrm{th}}} reference fluidized bed boiler. These scaling criteria are not used in the present radar-inversion analysis, for which detailed fluid-dynamic interpretation is outside the scope.} The radar-derived concentration profiles are consistent with pressure-based estimates across superficial gas velocities that produce distinct axial solids distributions and total solids holdup in the riser.

The remainder of the paper is organized as follows. \cref{Section:Theory} presents the theoretical framework, including the ensemble-averaged scattering formulation, signal processing, radar power model, beam characterization, and external calibration. \cref{Section:ExpSetupMaterials} describes the experimental setup, measurements, operating conditions, and material properties. \cref{Section:Methodology} presents the stabilized inversion method and the far-range boundary condition. \cref{Section:Results} presents the results, and \cref{Section:Conclusions} concludes the paper.

\section{Theory} \label{Section:Theory}
\subsection{Electromagnetic Scattering from a Single Sphere}
The interaction of a plane electromagnetic wave with an isolated scatterer is governed by classical electromagnetic scattering theory. In the free-space far-field regime, the scattered electric field $\mathbf{E}_s$ is related to the incident field $\mathbf{E}_i$ via the complex-valued scattering matrix $\mathbf{S}$ \cite{AbsScatBySmallParticles_CH4}:
\begin{equation}
\mathbf{E}_s = \frac{e^{-j2\pi f R / c}}{R} \, \mathbf{S}\,\mathbf{E}_i,
\label{eq:scattering_formalism}
\end{equation}
where $f$ is the frequency of the incident wave, $c$ is the speed of light in vacuum, and $R$ is the distance from the scatterer to the observation point. In a monostatic configuration, the backscattered signal strength is characterized by the radar cross-section (RCS) $\sigma_b$, which quantifies the effective backscattering area of the object. 

For a perfectly conducting or homogeneous dielectric sphere, rotational symmetry implies that the scattering matrix is diagonal and has equal diagonal elements, reducing the matrix formulation to a single complex-valued scattering amplitude $S$, for which \cite{book_PolarimetricDopplerWeatherRadar2001}:
\begin{equation}
\sigma_b = 4\pi \, |S|^2.
\label{eq:rcs_sphere}
\end{equation}
Scattering from a spherical particle of diameter $D$ and complex refractive index $\tilde{m}(f)$ is described by Mie theory, which provides an exact series solution. The cross-section $\sigma_\varphi$ is expressed in terms of the geometric cross-section and a dimensionless efficiency factor $Q_\varphi$ \cite{AbsScatBySmallParticles_CH5}:
\begin{equation}
\sigma_{\varphi}(f,D)=\frac{\pi D^2}{4}\,Q_{\varphi}\big(x(f,D),\tilde{m}(f)\big),
\label{eq:rcs_mie}
\end{equation}
where the size parameter $x(f,D) = \pi D / \lambda$ compares the particle size to the free-space wavelength $\lambda=c/f$. The subscript $\varphi \in \{b, s, a\}$ denotes backscattering, total scattering, and absorption cross-sections, respectively. The extinction cross-section is defined as $\sigma_e = \sigma_s + \sigma_a$, representing the total energy removed from the incident wave. The angular distribution of the scattered intensity is characterized by the asymmetry parameter (scattering anisotropy):
\begin{equation}
    g = \langle \cos\theta_s \rangle
      = \frac{\displaystyle\int_{4\pi} p(\theta_s)\cos\theta_s \,\mathrm{d}\Omega}
             {\displaystyle\int_{4\pi} p(\theta_s)\,\mathrm{d}\Omega},
    \label{eq:asymmetry_parameter}
\end{equation}
where $p(\theta_s)$ is the scattering phase function and $\theta_s$ is the scattering angle relative to the propagation direction of the incident wave \cite{prahl_miepython_2026,AbsScatBySmallParticles_CH4,AbsScatBySmallParticles_CH5}.
Forward-dominated scattering yields $g>0$, isotropic scattering $g=0$, and backward-enhanced scattering $g<0$ \cite{AbsScatBySmallParticles_CH4,AbsScatBySmallParticles_CH5}. 

Conductive media exhibit frequency-dependent behavior, with skin depth scaling as $\delta \propto 1/\sqrt{f}$ and asymptotically approaching the response of a perfect electrical conductor at high frequencies. This dispersive behavior is commonly modeled using the Drude formulation:
\begin{equation}
\tilde{m}^2 = 1 - \frac{\omega_p^2}{\omega \,(\omega + j\gamma)},
\label{eq:drude}
\end{equation}
where $\omega = 2\pi f$ is the angular frequency, $\omega_p$ is the plasma frequency, and $\gamma$ is the electron damping coefficient; the latter two are empirical, material-specific parameters \cite{Ordal_83,RefractiveIndex_Thz_2018}.
The Drude model captures free-electron behavior in many metals at microwave and sub-terahertz frequencies but does not include interband transitions, whose relevance is material dependent. The large magnitude of the refractive index of metallic spheres can introduce numerical difficulties in Mie-series evaluation, since electrically small particles ($x < 1$) may still satisfy $|\tilde{m}x| \gg 1$. This condition can lead to instability in the recursive computation of logarithmic derivatives in the classical Mie formulation. Wiscombe’s continued-fraction algorithm is employed to stabilize the evaluation of Bessel-function ratios \cite{WiscombeAlgorithm}.

\subsection{Ensemble-Averaged Scattering Formulation}\label{sec:ensemble_averaging}
The single-particle description in \cref{eq:rcs_mie} applies to an isolated scatterer. For a volumetric target $\Delta V(R)$ at range $R$ containing a statistically large number of particles drawn from a known polydisperse number-based size distribution $f_{\mathrm{N}}(D)$, the received power is described by its ensemble-mean response. 

The bulk response is obtained under three approximations, each assessed a posteriori in \cref{Section:Results}:
\begin{enumerate}
\item[(i)] \emph{Local homogeneity}: within each scattering volume $\Delta V(R)$ the particle number density $n(R)$ is uniform and particle positions are random.
\item[(ii)] \emph{Independent scattering}: the suspension is dilute ($\varepsilon_s \ll 1$) and the particles are electrically small ($x = \pi D / \lambda < 1$), so that each particle scatters as an isolated Mie scatterer and inter-particle (dependent) near-field coupling is neglected.
\item[(iii)] \emph{Single scattering}: the range-resolved return is formed by first-order backscatter, and radiation reaching the receiver after more than one scattering event is neglected; extinction of the directly propagating field is retained through the two-way attenuation factor introduced in \cref{RadarEquationSection}.
\end{enumerate}
Under conditions (i)-(iii), the particles occupy random positions within a scattering volume spanning many wavelengths, such that their scattered fields carry effectively uncorrelated relative phases and add incoherently in power. 

Ensemble averaging of the received power then suppresses inter-particle cross terms \cite{UlabyBook2014,AttenuationEffects_2011,Crane1971}, and the volumetric scattering response for a realization containing $N_p$ particles in $\Delta V(R)$, with $\mathbb{E}[N_p]=n(R)\,\Delta V(R)$, may be approximated as:
\begin{equation} 
\frac{1}{\Delta V(R)}\sum_{\ell=1}^{N_p}\sigma_{\varphi,\ell} \approx n(R)\,\langle\sigma_{\varphi}\rangle_{f,D}. 
\label{IncoherentScatteringEquation}
\end{equation}
The ensemble-averaged cross-section $\langle \sigma_{\varphi} \rangle_{f,D}$ characterizes the bulk scattering response per particle across the swept bandwidth \cite{THzRadarFrequencyAvgRCS_2010}. Scattering from a polydisperse ensemble is obtained by sequential averaging over frequency and particle size. First, the frequency-dependent cross-section is averaged across the bandwidth $B$ centered at $f_c$:
\begin{equation}
\langle \sigma_{\varphi}(D) \rangle_f = \frac{1}{B} \int_{f_c - B/2}^{f_c + B/2} \sigma_{\varphi}(f, D) \, \mathrm{d}f,
\label{eq:sigma_f_avg}
\end{equation}
accounting for the finite swept bandwidth of the FMCW waveform. Second, this quantity is averaged over the number-based particle-size distribution, $f_{\mathrm{N}}(D)$:
\begin{equation}
\langle \sigma_{\varphi} \rangle_{f,D}
= \int_{D_{\min}}^{D_{\max}} f_{\mathrm{N}}(D)\,\langle \sigma_{\varphi}(D) \rangle_f \, \mathrm{d}D,
\label{eq:sigma_fD_avg}
\end{equation}
where $f_{\mathrm{N}}(D)$ is normalized such that $\int f_{\mathrm{N}}(D)\,\mathrm{d}D = 1$. In the Rayleigh regime ($x \ll 1$), $\sigma_{\varphi}(f,D)\propto D^{6}$, implying that $\langle \sigma_{\varphi}\rangle_{f,D}$ is dominated by the upper tail of $f_{\mathrm{N}}(D)$. In this work, no Rayleigh approximation is invoked; instead, $\sigma_{\varphi}(f,D)$ is evaluated for each particle size via Mie theory \cite{THzRadarFrequencyAvgRCS_2010}. The asymmetry parameter in \cref{eq:asymmetry_parameter} is ensemble-averaged over $f_{\mathrm{N}}(D)$ in the same manner, weighted by the scattering cross-section, i.e., $\langle g \rangle_D = \langle g\,\sigma_s \rangle_D / \langle \sigma_s \rangle_D$. Hereafter, $g$ denotes this scattering-weighted ensemble value unless otherwise stated.

\begin{figure*}[!t]
\centering
\subfloat[]{\includegraphics[width=3in]{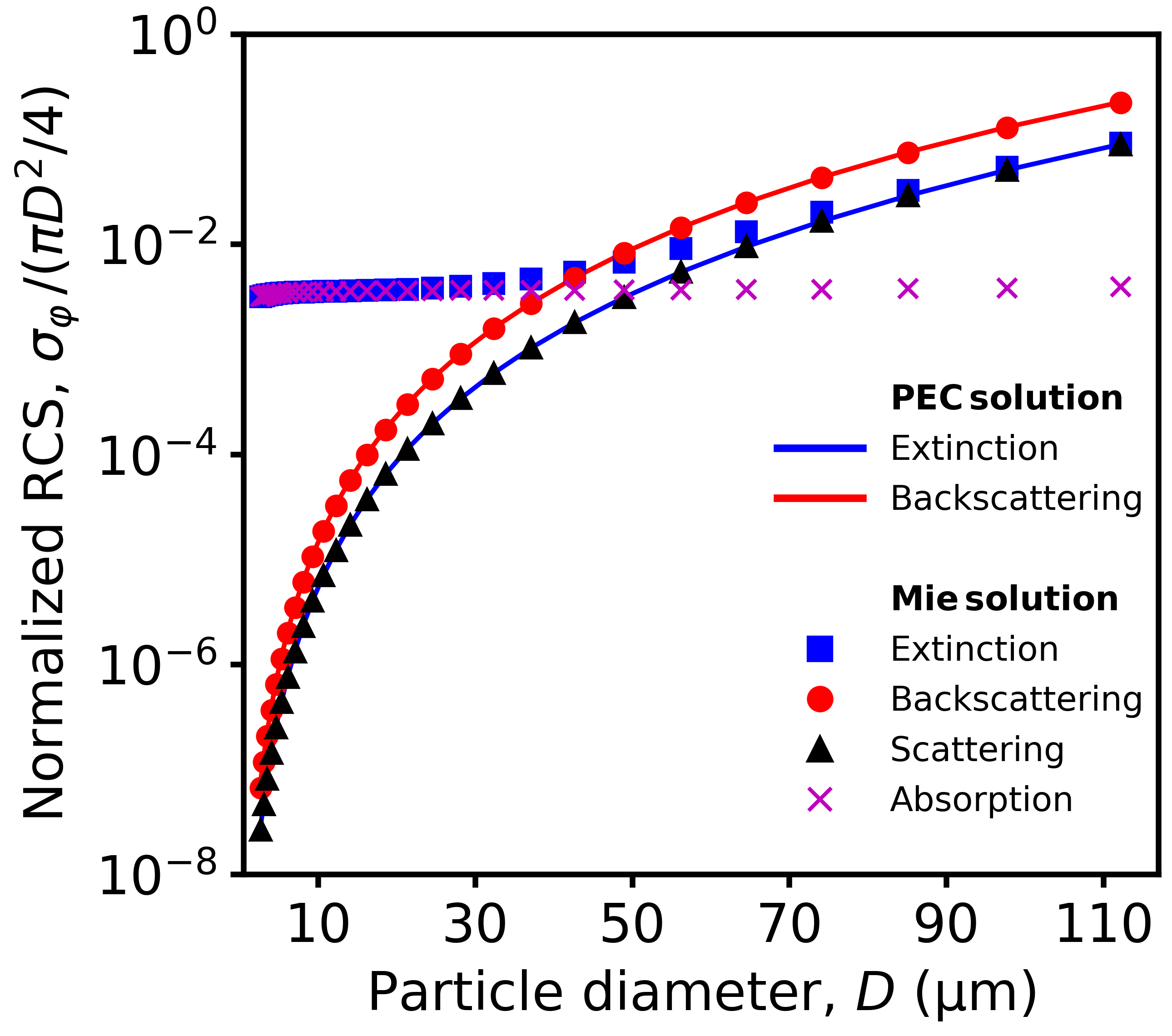}%
\label{fig:sigma_rayleigh}}
\hfil
\subfloat[]{\includegraphics[width=3in]{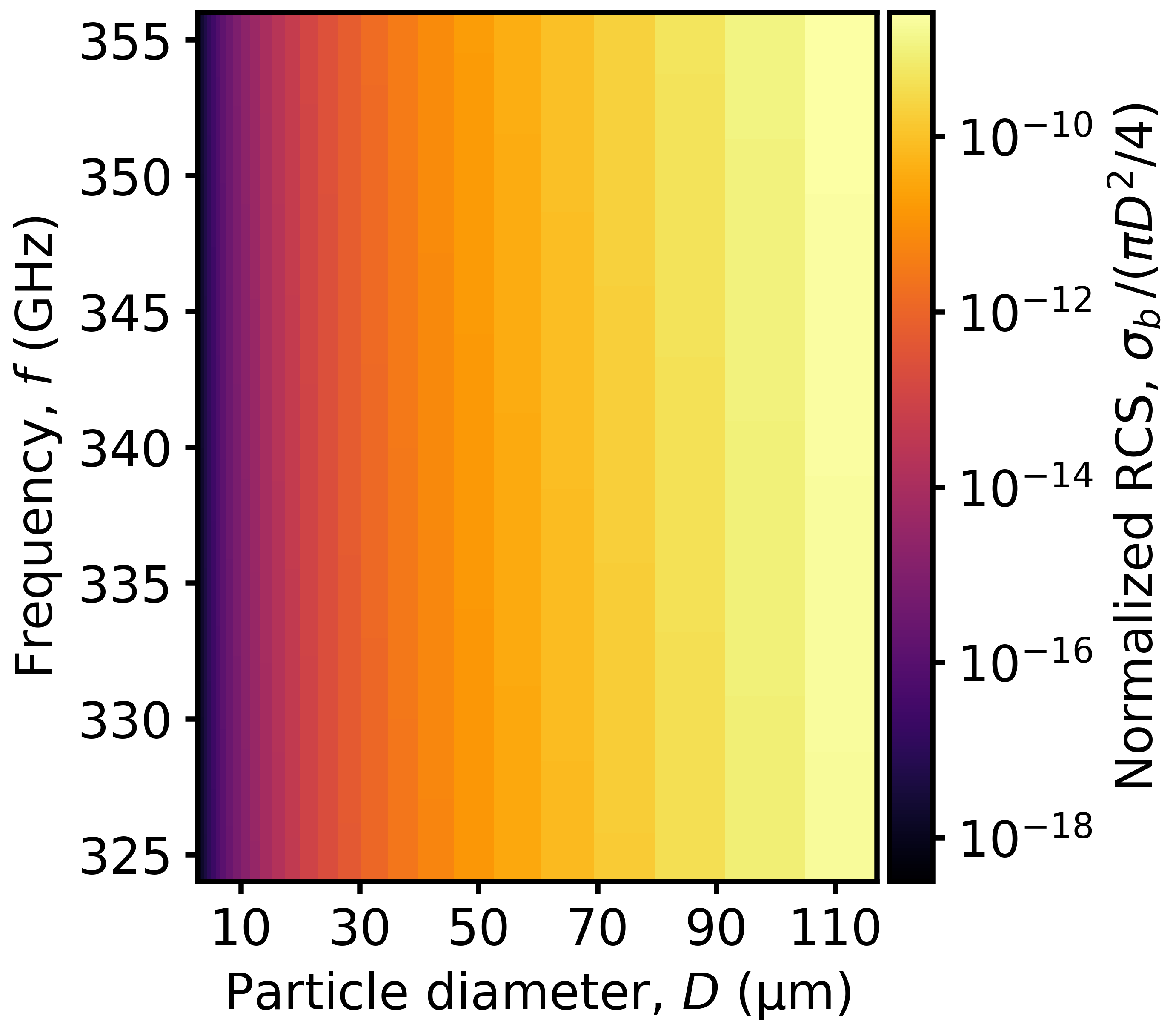}%
\label{fig:HeatMapRCS}}
\caption{(a) Frequency-averaged RCS computed for copper spheres over particle diameters $D\in\SIrange{2}{112}{\micro\meter}$, with the Rayleigh approximation included for comparison. (b) Normalized backscattering RCS as a function of particle diameter and frequency over the effective processed bandwidth $f\in\SIrange{325}{355}{\giga\hertz}$.}
\label{fig:rcs_pair}
\end{figure*}

\cref{fig:sigma_rayleigh} compares the frequency-averaged Mie RCS, computed over the effective processed bandwidth $f\in\SIrange{325}{355}{\giga\hertz}$ using the Wiscombe continued-fraction method, with the corresponding analytical Rayleigh solution for a perfectly electrically conducting (PEC) sphere \cite{RCSfromMETALsphere_microwave_to_optical_frequency_2014}. For metallic (copper) spheres spanning particle diameters $D\in\SIrange{2}{112}{\micro\meter}$, the size parameter satisfies $x<1$, while the condition $|\tilde{m}x|\gg1$ is maintained, consistent with the electrically small metallic limit. Agreement between the Mie and Rayleigh solutions over the particle-diameter interval considered supports numerically stable evaluation of the Mie series. Material-specific parameters used in the Drude model to evaluate the complex refractive index of copper at varying frequencies were taken from \cite{Ordal_83}. Potential surface oxidation of the copper powder is not quantified and may introduce uncertainty in the effective dielectric function used to represent the particles and, thus, in the associated single-particle RCS.

In \cref{fig:HeatMapRCS}, the normalized backscattering RCS $\sigma_{b}(f,D)$ is shown as a function of frequency and particle diameter. The variation of $\sigma_{b}(f,D)$ is primarily governed by particle size, with weaker frequency variation. This motivates a narrowband radar signal approximation in which $\sigma_\varphi(f,D)$ is treated as constant within each chirp. Therefore, the intraband dependence is evaluated at the center frequency and the ensemble averages reduce to particle-size averages, $\langle\sigma_{\varphi}\rangle_{f,D}\approx\langle\sigma_{\varphi}\rangle_{f_c,D}\equiv\langle\sigma_{\varphi}(f_c,D)\rangle_{D}$; hereafter, $\langle\cdot\rangle$ is omitted and $\sigma_{\varphi}$ denotes $\langle\sigma_{\varphi}(f_c,D)\rangle_{D}$.

\subsection{FMCW Pulse--Doppler Radar}\label{section:FMCW}
\begin{figure}[!t]
\centering
\includegraphics[width=3in]{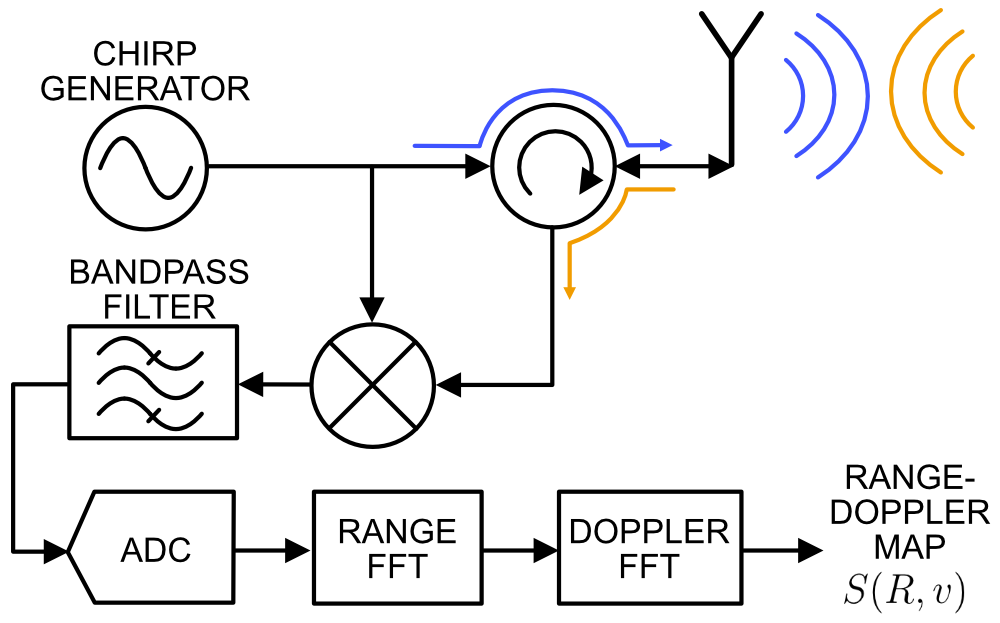}
\caption{Simplified functional block diagram of the radar system and digital signal processing.}
\label{fig:radar_block_diag}
\end{figure}

A frequency-modulated continuous-wave (FMCW) pulse--Doppler radar transmits a sequence of frequency-modulated (FM) pulses, commonly referred to as chirps. The transmitted linear chirp is $s_t(t_f,q)=e^{j\phi(t)}$, where the phase $\phi(t)=2\pi f_ct+\pi\mu t_f^2$. Here we distinguish between absolute time $t$ and fast time $t_f$, which repeats every pulse, i.e., $t=t_f+q \cdot PRI$, where $q$ is the pulse index and $PRI$ is the pulse repetition interval \cite{Carrara1995SpotlightSAR}. The chirp duration $T_c$ extends from $t_f=-T_c/2$ to $t_f=T_c/2$. The instantaneous frequency within a pulse is then \cite{mmWaveRadarFMCW2021}:
\begin{equation}
    f(t_f)=\frac{1}{2\pi}\frac{d\phi(t)}{dt_f}=f_c+\mu t_f
    \label{eq:instfreq}
\end{equation}
Here, $f_c$ denotes the carrier frequency and $\mu = B/T_c$ is the chirp rate, determined by the total swept bandwidth $B$. \cref{eq:instfreq} assumes a large time-bandwidth product ($T_c B\gg 1$). For a moving point target located at range $R_t(t)$, the received signal is equal to a version of the transmitted pulse that has been scaled in amplitude and delayed in time by the two-way propagation $\tau = 2R_t(t)/c$. The received signal is modeled as:
\begin{equation}
s_r(t_f,q) = A \, s_t(t - \tau,q) = Ae^{j2\pi f_c(t-\tau)}e^{j\pi\mu (t_f-\tau)^2},
\end{equation}
where $A$ is a complex amplitude that accounts for system gains, propagation loss, and the target scattering characteristics $\sigma_{\varphi}$. The received chirp duration $T_c$ still extends approximately from $t_f=-T_c/2$ to $t_f=T_c/2$ since $T_c \gg \tau$ for FMCW radar. The received signal is then mixed with the transmitted signal to obtain the intermediate-frequency (IF) signal \cite{340GHzHomodyneTransceiver_2016}:
\begin{align}
s_{\mathrm{IF}}(t_f,q)&=s_t^{*}(t_f,q)\,s_r(t_f,q) \nonumber\\
&=A\,e^{-j2\pi \mu \tau t_f}\,e^{-j2\pi f_c \tau}\,e^{j\pi\mu \tau^2},
\label{x_IF}
\end{align}
The signal processing operations for range and velocity estimation are simplified under the following assumptions:
\begin{enumerate}
\item Over all repeated pulses, the target range $R_t(t)$ varies by much less than the range resolution $\Delta R = c/(2B)$. Under this condition, the first phase term in \cref{x_IF} is well approximated by $\textup{exp}(-j2\pi f_b t_f)$, where $f_b=\mu \Bar{\tau}$ is the beat frequency associated with an average range $\Bar{R}$ and time delay $\Bar{\tau}=2\Bar{R}/c$.
\item A stop-and-hop model is adopted, i.e., the target range is constant within each pulse and updates between pulses with constant radial velocity $v_t$ \cite{StopAndHop_2014}. The second phase term in \cref{x_IF} then reduces to $\textup{exp}(-j2\pi f_D q \cdot PRI)$, where $f_D=2v_tf_c/c$ is the Doppler frequency.
\item The third phase term in \cref{x_IF}, commonly termed the video phase, satisfies $\mu \tau \ll 2f_c$ and thus varies negligibly relative to the Doppler term. It is treated as a constant phase and neglected in the remainder of the analysis.
\end{enumerate}
Under assumptions 1--3, after sampling repeated chirps with an analog-to-digital converter (ADC) and neglecting constant phase terms, the digital IF signal in \cref{x_IF} reduces to:
\begin{equation}
s_{\mathrm{IF}}(pT_s,q)= A\,e^{-j2\pi f_{b} pT_s}\,e^{-j2\pi f_D q\mathrm{PRI}},
\label{x_IF_digital}
\end{equation}
where $p$ is the ADC sample index within a chirp and $T_s$ is the sampling interval \cite{SignalProcessing_2024}. From this representation, the range $\Bar{R}$ and radial velocity $v_t$ can be extracted from $s_{\mathrm{IF}}(p,q)$ via a two-dimensional discrete time Fourier transform (DTFT):
\begin{align}
S(R,v) &= \textup{DTFT}_{pT_s,q}\!\left\{ s_{\mathrm{IF}}(p,q)\right\} \nonumber\\
       &= A\cdot\mathrm{IRF}(R,v) \ast \delta\!\left(R-\Bar{R},\;v-v_t\right),
\end{align}
where $\mathrm{IRF}(R,v)$ is the impulse response function and $\delta \left (R, v \right )$ is the Dirac delta function. The impulse response function has the form of a two-dimensional periodic sinc function (Dirichlet kernel), implying that the target response folds (aliases) across the edges of the discretized range--Doppler matrix when the target exceeds the range or velocity limits. While bandpass filters avoid such folding in the range dimension, velocity folding occurs when $ \left | v_r\right | > c/(4f_c \mathrm{PRI})$. The periodic sinc functions in the range and velocity dimensions give rise to a finite main lobe width, defining the range and velocity resolutions, and sidelobes, which spread the target energy in each dimension. These sidelobes are suppressed using window functions (e.g., Hanning windows), which have been omitted in the above signal model for simplicity. The range resolution, $\Delta R = c/(2B)$, depends on the bandwidth of the transmitted signal and the velocity resolution, $\Delta v = c/(2f_c \mathrm{CPI})$, depends on the center frequency and coherent processing interval $\mathrm{CPI}=n_{\mathrm{PRI}} \cdot \mathrm{PRI}$ (the total time used to measure $n_{\mathrm{PRI}}$ pulses) \cite{Review_SignalProcessingMethods_2017}. The velocity information is not used in this study, but the DTFT in the velocity dimension is still performed as it provides a factor-of-$n_{\mathrm{PRI}}$ gain in signal-to-noise ratio (SNR). \cref{fig:radar_block_diag} shows a simplified block diagram of the radar measurement described above.

\begin{figure*}[!t]
  \centering
  \includegraphics[width=\textwidth]{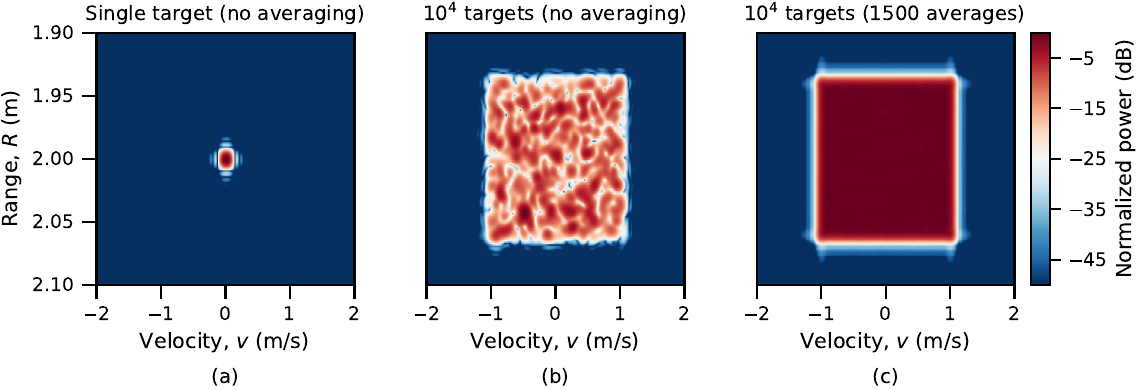}
  \caption{Simulated range--Doppler matrix power $\left|S(R,v)\right|^{2}$ for (a) a single stationary point target at $R=\SI{2}{\meter}$, (b) a random uniform distribution of $10^{4}$ point targets over $R \in [1.94,2.06]\,\mathrm{m}$ and $v \in [-1,1]\,\mathrm{m/s}$, and (c) the case in (b) after averaging 1500 independent realizations of $\left|S(R,v)\right|^{2}$.}
  \label{fig:fmcw_sim}
\end{figure*}  

\cref{fig:fmcw_sim}\textcolor{blue}{a} shows a simulated range--Doppler power matrix, $\left | S(R,v)\right |^2$, in decibels (dB). The radar response of a stationary point target (e.g., a small metallic sphere) at $R=2~\mathrm{m}$ is shown. This 2D response is defined by the impulse response $10 \, \textup{log}_{10}\left | \mathrm{IRF}(R,v)\right |^2$, where Hanning windows have been included in the simulation. The main lobe defines the resolution cell, and the peak coincides with the target range and velocity. However, if multiple targets exist within a resolution cell, their returns add with random relative phases. This is illustrated in \cref{fig:fmcw_sim}\textcolor{blue}{b}, where $10^{4}$ targets are uniformly distributed over $R\in[1.94,2.06]~\mathrm{m}$ and $v\in[-1,1]~\mathrm{m/s}$. The resulting speckle pattern has power fluctuations that follow an exponential distribution. To reduce the variance of the power estimate, the range--Doppler measurement is repeated and the power averaged. The result after 1500 averages is shown in \cref{fig:fmcw_sim}\textcolor{blue}{c}, where the power becomes approximately constant due to the uniform target density.

Given that velocity information is not used in this study, the range--Doppler power spectrum is integrated over all Doppler bins, yielding an estimate of the received power as a function of range:
\begin{equation}
\widehat{P}_r(R)=K(R)\sum_{\kappa}\sum_{v}\left| S^{\kappa}(R,v) \right|^2,
\end{equation}
where $K(R)$ aggregates system-dependent constants and is determined through system calibration. The index $\kappa$ denotes the repeated range--Doppler measurement. The solids concentration is subsequently estimated from the measured $\widehat{P}_r(R)$ using the power model introduced in the following section.

\subsection{Radar Equation for Volume Scattering}\label{RadarEquationSection}
We consider a monostatic radar system operating in a spherical coordinate framework $(r, \theta, \phi)$, with the origin defined at the antenna phase center. Here, $r$ denotes the radial distance from the antenna, $\theta$ is the elevation angle relative to the boresight, and $\phi$ is the azimuthal angle.

Under the assumption of a narrow well-collimated beam, the mean received power $P_r(R,v)$ from targets with velocity $v$ within an infinitesimal differential volume element $\mathrm{d}V = r^{2} \sin\theta\, \mathrm{d}\theta\, \mathrm{d}\phi \, \mathrm{d}r$ can be expressed as \cite{VolumeBasedRadarEquation_Dovia_1979}:
\begin{multline}
 P_r(R,v) = \frac{P_t \lambda^2 G_0^2}{(4\pi)^3}
\iiint_{V} \left | \mathrm{IRF}(r-R,v)\right |^2 \, \frac{g_{\mathrm{n}}^2(r, \theta, \phi)}{r^4} \\
\, \eta(r) \, e^{-2\alpha(r)} \, \mathrm{d}V + P_0(R).
\label{RadarEquationTime}
\end{multline}
In \cref{RadarEquationTime} $P_t$ is the peak transmitted power, $\lambda$ is the wavelength at the center frequency, $G_0$ is the on-axis antenna gain, $g_{\mathrm{n}}(r,\theta,\phi)$ is the normalized antenna power pattern, and $\eta(r)$ represents the volumetric backscattering coefficient. The range-dependent background noise power $P_0$ is subtracted after a reference measurement with an empty riser and is, therefore, omitted in expressions of measured power. The exponential attenuation term incorporates the path-integrated extinction:
\begin{equation}
\alpha(R) = \int_0^R k(r)\, \mathrm{d}r,
\end{equation}
where $k(r)$ is the local volumetric extinction coefficient, accounting for absorption and scattering losses. From the assumption of a well-collimated beam, lateral variations in $\eta$ and $k$ across the illuminated cross-section are negligible, i.e., $\eta(r,\theta,\phi)\approx\eta(r)$ and $k(r,\theta,\phi)\approx k(r)$.

If the range and velocity resolutions are small relative to the spatial and velocity scales over which the integrand in \cref{RadarEquationTime} varies, $\left | \mathrm{IRF}(r-R,v)\right |^2$ can be approximated as a Dirac delta function in range. The convolution integral in \cref{RadarEquationTime} then simplifies to:
\begin{equation}
P_r(R) = \frac{K(R)}{R^4} \, \eta(R) \, e^{-2\alpha(R)} \, \Delta V(R).
\label{SimplifiedRadarEquation}
\end{equation}
In \cref{SimplifiedRadarEquation}, the effective scattering volume $\Delta V(R)$ incorporates the directional antenna response as:
\begin{equation*}
    \Delta V(R) = \int_{R - \Delta R / 2}^{R + \Delta R / 2} \int_0^{2\pi} \int_0^{\pi} p_{\mathrm{ant}}^2(r, \theta, \phi) \, r^2 \sin\theta \, \mathrm{d}\theta \, \mathrm{d}\phi \, \mathrm{d}r.
\end{equation*}
In this form, \cref{SimplifiedRadarEquation} is generalized such that $\Delta V(R)$ and $K(R)$ can be obtained from established analytical expressions \cite{ProbertJones1962} or, alternatively, from empirical measurements and external calibration procedures \cite{RCS_and_calibration_2023}. Regardless of this closure, \cref{SimplifiedRadarEquation} links the return power $P_r(R)$ to the range-dependent volumetric coefficients $\eta(R)$ (backscatter) and $k(R)$ (extinction). The factor $e^{-2\alpha(R)}$ couples $P_r(R)$ to all preceding ranges, so recovering $\eta$ and $k$ from $P_r$ is a nonlinear inverse problem. In strongly attenuating media, this inversion can be severely ill-conditioned: small perturbations in $P_r$ or the calibration parameters can induce large errors in the recovered concentration profiles.

Under the incoherent scattering approximation introduced in \cref{sec:ensemble_averaging}, the radar-observable volumetric coefficients for the medium at range $R$ are given by:
\begin{subequations}
\begin{align}
\eta(R) &= n(R)\,\sigma_{b}\, \label{eq:scattering_eta}\\
k(R) &= n(R)\,\sigma_{e}. \label{eq:scattering_k}
\end{align}
\end{subequations}
Here, $\sigma_{b}$ and $\sigma_{e}$ are particle-size-averaged at $f_c$, i.e., $\sigma_{\varphi}\equiv\langle\sigma_{\varphi}(f_c,D)\rangle_{D}$ with $\varphi\in\{b,e\}$. The particle number density $n(R)$ relates to the solids volume fraction $\varepsilon_s(R)$ through the mean particle volume:
\begin{equation}
\varepsilon_s(R)=n(R)\left\langle V_p \right\rangle
= n(R)\,\frac{\pi}{6}\int_{D_{\min}}^{D_{\max}} D^3 f_{\mathrm{N}}(D)\,\mathrm{d}D.
\label{eq:solids_volume_concentration}
\end{equation}
If $n(R)$ is known and satisfies assumption (i) in \cref{sec:ensemble_averaging}, the mean nearest-neighbor distance is approximated as \cite{Bansal_1972}:
\begin{equation}
\langle d \rangle \approx 0.554 \, n^{-1/3}(R).
\label{eq:mean_distance}
\end{equation}
The prefactor 0.554 corresponds to the expected nearest-neighbor spacing for a three-dimensional spatially random point distribution. The ratio $\langle d \rangle / \lambda$ serves as a diagnostic of dependent (near-field) inter-particle coupling. Such effects can occur for $\langle d \rangle \lesssim \lambda$, whereas the independent-scattering closure of \cref{eq:scattering_eta,eq:scattering_k} is expected to become more accurate as $\langle d \rangle / \lambda$ increases. The ratio is evaluated a posteriori in \cref{Section:Results}.

\subsection{Beam Propagation and Spatial Characterization}
Originally developed for laser optics to characterize the propagation of elliptical Gaussian beams, the ISO~11146 method \cite{iso2021part1,iso2021part2,iso2004part3} provides a statistical framework for moment-based evaluation of the $1/e^2$ beam widths from a measured irradiance distribution. Here, the standard is adapted to estimate 3-dB half-power beam radii of the radar antenna.

For discrete irradiance measurements across the transverse plane $E_{ij}(R) = E(x_i, y_j, R)$, the power-weighted centroid is computed as:
\begin{equation*}
\mathbf{m}(R) = \frac{\sum_{i,j} \mathbf{X}_{ij} E_{ij}(R)}{\sum_{i,j} E_{ij}(R)}, \quad \mathbf{X}_{ij} = [x_i, y_j]^{\mathsf{T}},
\end{equation*}
and the beam extent is described by the covariance matrix:
\begin{equation*}
\mathbf{M}(R) = \frac{\sum_{i,j} \left(\mathbf{X}_{ij} - \mathbf{m}(R)\right)\left(\mathbf{X}_{ij} - \mathbf{m}(R)\right)^{\mathsf{T}} E_{ij}(R)}{\sum_{i,j} E_{ij}(R)}.
\end{equation*}
Its eigenvalues $\Lambda_x(R)$ and $\Lambda_y(R)$ correspond to the elliptical beam’s principal axes. Let $w_\xi(R)$ denote the half-power beam radius, i.e., the semi-axis of the 3-dB contour, along principal axis $\xi\in\{x,y\}$. For quasi-Gaussian profiles:
\begin{equation}
\label{eq:w3dB}
 w_\xi(R) =  \sqrt{2\ln 2}\,\sqrt{\Lambda_\xi(R)}, \quad \xi \in \{x,y\}.
\end{equation}
The 3-dB cross-sectional area is then given by $\mathcal{A}(R) = \pi w_x(R) w_y(R)$ and incorporated in the description of the effective scattering volume as:
\begin{align*}
\Delta V(R) \approx \mathcal{A}(R) \,\Delta R = \pi \,w_x(R)\, w_y(R) \,\frac{c}{2B}.
\end{align*}

\subsection{External Target Calibration} \label{sec:theory_calibration}
External calibration procedures commonly use reference targets of known RCS to characterize the radar's response. A large, perfectly conducting sphere---in the geometric-optics regime ($D_{\mathrm{sph}}\gg\lambda$)---is typically chosen as the reference because of its polarization-independent backscatter and analytically known RCS, $\sigma_{\mathrm{sph}}=\pi(D_{\mathrm{sph}}/2)^2$ \cite{IEEE_standardPractice_2020,RCS_and_calibration_2023}. For such a target, the radar equation (\cref{SimplifiedRadarEquation}) simplifies to $P_{\mathrm{peak}}(R)=K(R)\sigma_{\mathrm{sph}}/R^4$, where $P_{\mathrm{peak}}(R)$ is the peak power at range $R$ and $K(R)$ is the system calibration factor, encompassing the two-way antenna gain pattern, transmitted power, receiver sensitivity, signal-processing gains, and other losses. By measuring $P_{\mathrm{peak}}(R)$ for a target with known $\sigma_{\mathrm{sph}}$, $K(R)$ can be determined empirically \cite{Skolnik_RadarBook}. In this work, calibrations are performed at multiple distances in both the near and far fields, allowing $K(R)$ to capture range-dependent antenna and propagation effects.

\section{Experimental Setup and Materials} \label{Section:ExpSetupMaterials}
\subsection{Radar System}\label{RadarSystemDescription}
The \SI{340}{\giga\hertz} FMCW range--Doppler radar uses a frequency-upconverted and frequency-multiplied architecture. A chirp with a \SI{1}{\giga\hertz} bandwidth, centered at \SI{1}{\giga\hertz}, is generated by an FPGA-controlled arbitrary waveform generator operating at \SI{4}{\giga\sample\per\second} and supports arbitrary pulse trains within a CPI\@. The chirp is up-converted to the X-band using a \SI{9.6}{\giga\hertz} local oscillator and applied to the transceiver. The X-band chirp is then multiplied to \SI{340}{\giga\hertz} by a three-stage chain consisting of an active $\times 8$ InGaAs pHEMT frequency multiplier, a $\times 2$ Schottky diode frequency doubler, and a final $\times 2$ Schottky front-end circuit that also operates as a subharmonic mixer, giving an overall $\times 32$ multiplication and an effective \SI{32}{\giga\hertz} bandwidth. The subharmonic-mixer architecture supports simultaneous transmission and reception. The radiated signal is collimated by a horn-fed off-axis parabolic mirror. Received echoes are mixed with the transmit reference, down-converted to baseband, and digitized by a \SI{250}{\mega\sample\per\second}, 14-bit ADC\@. All system clocks share a common \SI{10}{\mega\hertz} reference, providing coherence across the transmitter and receiver.

The physical bandpass filter, as shown in the block diagram in \cref{fig:radar_block_diag}, has lower and upper cutoff frequencies of
\SI{600}{\kilo\hertz} and \SI{20}{\mega\hertz}, respectively. Consequently, the upper cutoff attenuates signals corresponding to target ranges greater than
approximately \SI{3.84}{\meter}. In addition, a digital high-pass filter with a soft transition is applied to the beat-frequency spectrum during signal
processing, with its roll-off beginning near \SI{6}{\mega\hertz}, corresponding to $R\approx\SI{1.15}{\meter}$. The corresponding range-dependent response is included in the system calibration $K(R)$, since identical filtering and processing are applied to the calibration and measurement data.

The parameters in \cref{tab:radar_parameters} define the unambiguous range and velocity limits, $R_{\max}$ and $v_{\max}$, and the corresponding resolutions \cite{RadarBased_Tomas_2023_IEEE}. In the selected configuration, linear up-chirps are emitted across $B$, yielding $\Delta R\approx\SI{5}{\milli\meter}$ and $R_{\max}=\SI{6}{\meter}$, which exceeds the \SIrange{1}{4}{\meter} range of interest in this study. The Doppler spectrum spans $\pm\SI{4.3}{\meter\per\second}$ with resolution $\Delta v=\SI{0.067}{\meter\per\second}$.
\begin{table}[!t]
    \centering
    \caption{Radar and pulse design parameters used in this work.}
    \label{tab:radar_parameters}
    \begin{tabular}{lll}
        \toprule
        \textbf{Parameter} & \textbf{Symbol} & \textbf{Value} \\
        \midrule
        Center frequency & $f_c$ & \SI{340}{\giga\hertz} \\
        Pulse bandwidth & $B$ & \SI{32}{\giga\hertz} \\
        Chirp rate & $\mu$ & \SI{780.5}{\mega\hertz\per\micro\second} \\
        Chirp duration & $T_c$ & \SI{41}{\micro\second} \\
        Pulse repetition interval & $\mathrm{PRI}$ & \SI{51.2}{\micro\second} \\
        Pulses coherently processed & $n_{\mathrm{PRI}}$& 128 \\
        \bottomrule
    \end{tabular}
\end{table}
A custom-designed integrated smooth-walled spline-profile horn antenna is used for transmission and reception, providing spatial and phase alignment \cite{AntennaHorn340GHzRadar_2023}. Beam collimation is achieved downstream of the horn antenna via an off-axis mirror configuration comprising a planar offset mirror and a parabolic gold mirror with a projected aperture of \SI{10}{\centi\meter}. The parabolic mirror is the last effective aperture before free-space propagation and is designed to reduce the wavefront curvature of the transmitted field (quasi-collimation).

Temporal windowing isolates individual chirps for (i) acquisition of backscatter profiles via analog-to-digital conversion and (ii) Doppler-resolved velocity estimation through coherent integration (FFT) over $n_{\mathrm{PRI}}$ chirps. The time between successive pulses is defined by the pulse repetition interval $\mathrm{PRI}$, resulting in a $\mathrm{CPI}$ of \SI{6.6}{\milli\second}. Accounting for processing overhead, the system produces a single range--Doppler matrix at an output rate of \SI{5}{\hertz}. For additional hardware and signal-processing details, see \cite{RadarBased_Tomas_2023_IEEE}.

\subsection{External Calibration and Beam Characterization Setup}
\label{sec:calibration_setup}
Before in situ deployment, a controlled laboratory procedure was established for (i) characterization of the beam geometry---width, divergence, and orientation---and (ii) determination of the range-dependent calibration coefficient. For this purpose, a reference target was raster-scanned in two-dimensional planes normal to the radar beam. The target was a metallic sphere with a diameter of \SI{16}{\milli\meter}. At \SI{0.34}{\tera\hertz}, the corresponding free-space wavelength is approximately \SI{0.88}{\milli\meter}; thus, $D_{\mathrm{sph}}\gg\lambda$, placing the sphere in the geometric-optics scattering regime \cite{RCSfromMETALsphere_microwave_to_optical_frequency_2014}. The scan comprised nine axial positions along the radar beam at ranges between \SI{0.5}{\meter} and \SI{3.9}{\meter}, covering both near- and far-field regions. At each position, irradiance distributions $E(x,y,R)$ were obtained over a $25\times25$ grid with a pixel size of $\SI{5}{\milli\meter}\times\SI{5}{\milli\meter}$. No active beam-alignment scheme was used because the electrically large sphere provides a polarization-independent, axially symmetric RCS and is comparatively insensitive to small angular or translational misalignments.


\subsection{Circulating Fluidized Bed Configuration}
\begin{figure*}[!t]
\centering
\includegraphics[width=0.75\textwidth]{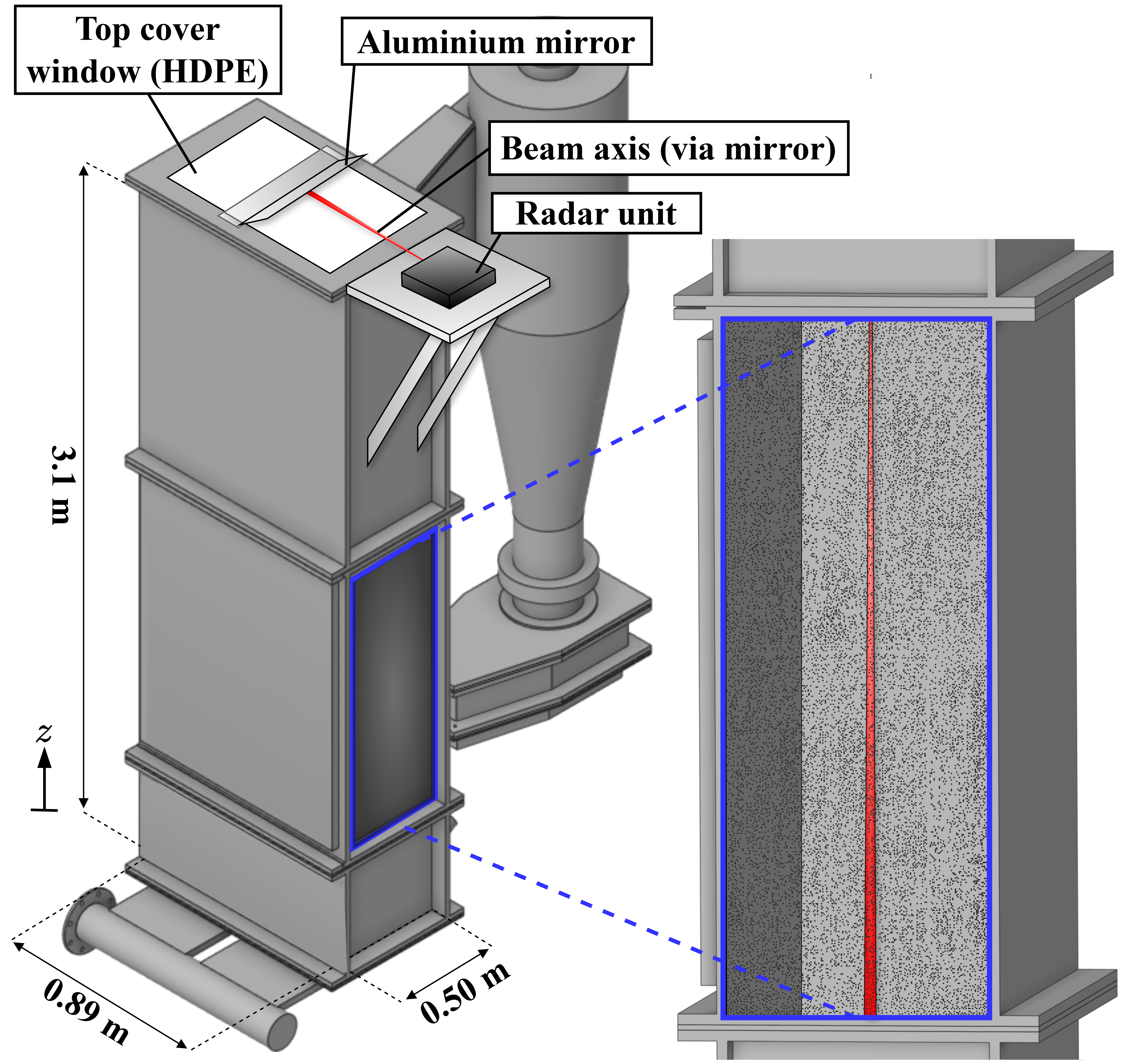}
\caption{Experimental setup and radar measurement geometry. The \SI{340}{\giga\hertz} radar is mounted above the riser and illuminates a vertical line-of-sight via an aluminum mirror through the HDPE roof. Inlet air enters through the bottom gas distributor.}
\label{fig:cfb_radar_setup}
\end{figure*}

The experimental campaign was conducted in the riser section of a CFB unit with interior dimensions of \SI{3.1}{\meter} (height)~$\times$~\SI{0.89}{\meter}~$\times$~\SI{0.50}{\meter}, as illustrated in \cref{fig:cfb_radar_setup}. The riser roof is made of high-density polyethylene (HDPE), forming a low-loss transmission window for the incident radar beam. 
The radar beam exits the antenna horizontally relative to the riser and is redirected vertically downward by a tilted aluminum mirror, yielding approximately normal incidence on the HDPE roof. The riser outlet is located on one of the long lateral walls, spanning approximately \SIrange{0.30}{0.60}{\meter} below the riser roof.

\begin{figure}[!t]
\centering
\includegraphics[width=3.0in]{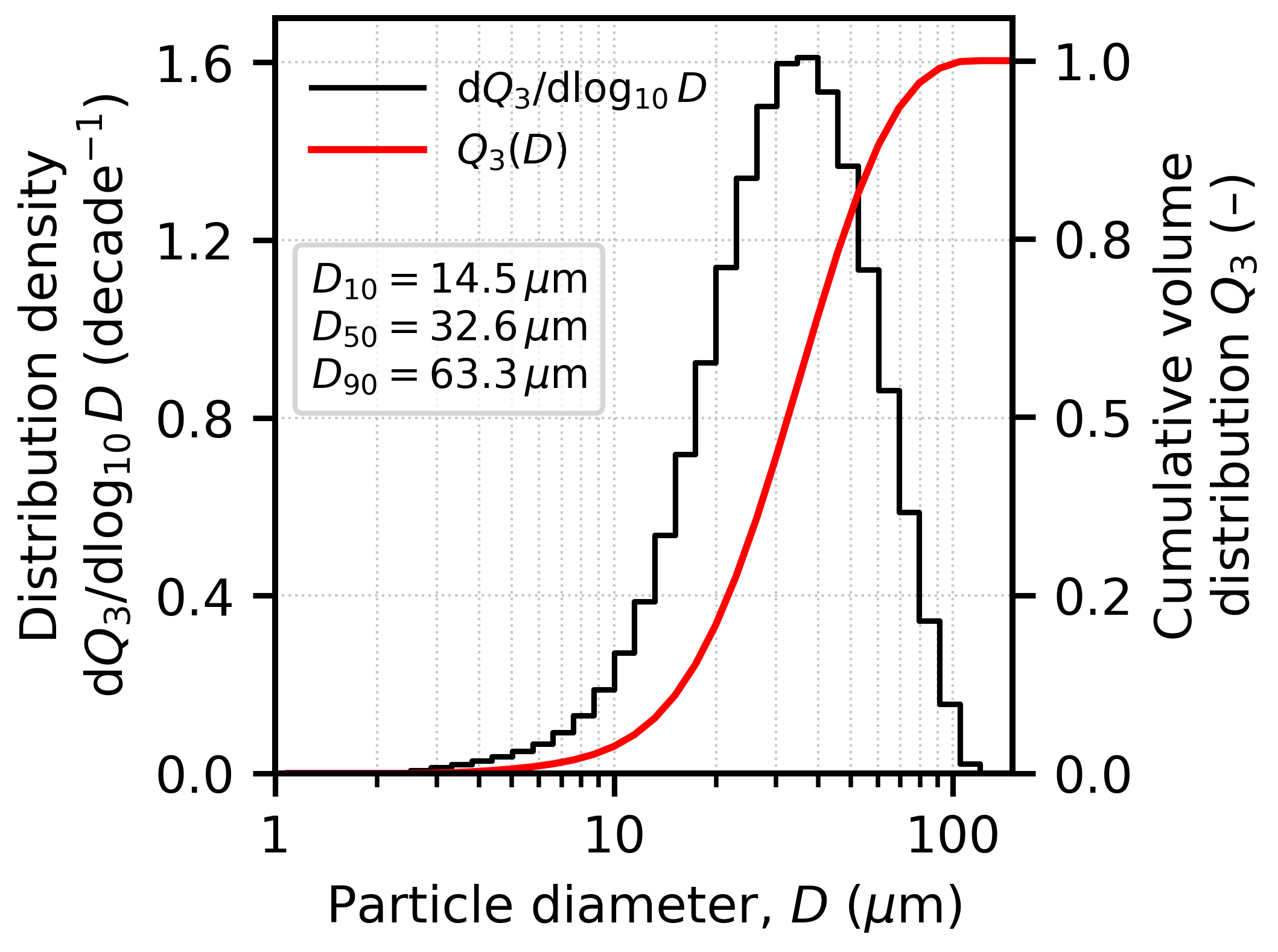}
\caption{Volume-based particle-size distribution of the copper powder measured by laser diffraction. The cumulative volume distribution $Q_3(D)$ and the corresponding log-density $\mathrm{d}Q_3(D)/\mathrm{d}\log_{10}(D)$ are shown, with the percentile diameters $D_{10}$, $D_{50}$, and $D_{90}$ indicated.}
\label{fig:psd}
\end{figure}

The bed material consisted of copper powder with density $\rho_s=\SI{8920}{\kilo\gram\per\cubic\meter}$ and volume-median diameter $D_{50}=\SI{32.6}{\micro\meter}$. The particle-size distribution was measured by laser diffraction (Malvern Mastersizer~2000) and is shown in \cref{fig:psd} as the cumulative volume distribution $Q_3(D)$ and the corresponding log-density $\mathrm{d}Q_3(D)/\mathrm{d}\log_{10}(D)$. For ensemble averaging of single-particle scattering properties, the volume-based distribution was converted to a number-based size distribution $f_{\mathrm{N}}(D)$ under the assumptions of spherical particles and uniform material density. 

The metallic powder was selected to satisfy Glicksman’s scaling laws for fluid-dynamic similarity with an industrial \SI{330}{\mega\watt_{\mathrm{th}}} CFB boiler \cite{GLICKSMAN1988,GLICKSMAN1993_simplifiedRelations}. The scaling relations are detailed in \cite{Tove2021_ScalingLaws}.

The riser is instrumented with 27 pressure transducers distributed along the height of one riser sidewall \cite{RadarBased_Carolina_2023_CFBs,Tove2021_ScalingLaws}. Consequently, the cross-sectional mean solids volume fractions $\varepsilon_s$ were estimated from differential-pressure measurements using the hydrostatic balance:
\begin{equation}
    \Delta p = \left( \rho_g(1-\varepsilon_s) + \rho_s\varepsilon_s \right)g_z\Delta z,
    \label{dP}
\end{equation}
where $\rho_g$ is the gas density, $\Delta z$ is the vertical spacing between adjacent transducers, and $g_z$ is the gravitational acceleration.

Industrial CFB boilers generally exhibit nearly uniform horizontal solids concentrations within the core, which spans 87--100\% of the cross section \cite{FluiDynamicBoundaryLayersCFBboilers1995,ToveHenrik_DNS}. The present facility is fluid-dynamically scaled from the \SI{330}{\mega\watt_{\mathrm{th}}} reference boiler according to the scaling relations in \cite{Tove2021_ScalingLaws}, supporting comparable large-scale solids-flow characteristics. Differential-pressure measurements provide established estimates of the cross-sectional mean solids concentration in CFB risers \cite{WERTHER1999_MeasurementInFluidizedBeds,ReviewMeasurementFluidizedBeds2008}. Together with the expected uniformity of solids concentration across the core region at a given height , the pressure-derived solids-volume-fraction profiles serve as reference estimates for evaluating the radar retrieval in terms of concentration levels and axial trends under the present conditions, although over substantially coarser axial averaging intervals than the \SI{5}{\milli\meter} radar range resolution.


\subsection{Riser Measurement Reference System}\label{Riser_reference_system}
The radar is mounted above the riser and views vertically downward into the riser via an aluminum mirror, as illustrated in \cref{fig:cfb_radar_setup}. Hence, the radar range $R$ increases with
the distance from the roof. For simpler interpretation and comparison to the pressure-based estimates, the range $R$ is expressed in the riser-relative axial coordinate as:
\begin{equation}
  z = R_{\mathrm{ref}} - R,
  \label{eq:z_coordinate}
\end{equation} 
where $R_{\mathrm{ref}} = \SI{4.115}{\meter}$ places the $z=0$ reference level approximately \SI{0.05}{\meter} below the distributor-plate surface, with the associated double-reflection feature extending from approximately $z=\SI{0.05}{\meter}$ to $z=\SI{0.15}{\meter}$. Increasing $z$ therefore denotes increasing axial height in the riser. The HDPE roof maps to $z\approx\SI{3.15}{\meter}$. This transformation is used to report the radar-based profiles in \cref{Section:Results}. Consequently, the Doppler-velocity axis follows the convention: $v_z>0$ denotes upward motion (toward the HDPE roof) and $v_z<0$ downward motion (toward the distributor plate).

\subsection{Operating Conditions and Data Acquisition} \label{sec:data_acquisition}
Air is introduced to the CFB riser through a bottom perforated plate that distributes the gas flow and transfers momentum to the solids, producing a gas--solid suspension in the riser. This study applied superficial gas velocities $U_g$ of \SI{0.66}{\meter\per\second}, \SI{0.79}{\meter\per\second}, and \SI{1.10}{\meter\per\second}, producing three distinct particle distributions and solids-holdup levels within the riser. The latter was indirectly quantified by the total pressure drop, measured between pressure taps \SI{0.016}{\meter} above the gas distributor and \SI{0.36}{\meter} below the top cover, which was \SI{0.59}{\kilo\pascal}, \SI{0.56}{\kilo\pascal}, and \SI{0.47}{\kilo\pascal}, respectively. For each operating condition, radar and pressure data were acquired simultaneously for \SI{5}{\minute} to ensure robust statistics (the dynamics of the flow show a main frequency 2-5 Hz).

As described in \cref{section:FMCW}, the present study considers signal power integrated over the velocity domain, and Doppler information is not used. The range-dependent power used for inversion was obtained by incoherently accumulating $\lvert S^{\kappa}(R,v)\rvert^2$ over Doppler bins and subsequently averaging the repeated CPIs acquired during \SI{5}{\minute}, yielding 1500 coherently processed range--Doppler matrices. Apart from empty-riser background subtraction, the only additional post-processing step applied to the resulting incoherent range profile was linear interpolation across the zero-Doppler bin to mitigate stationary contributions (static clutter and DC/leakage terms). The background signal was obtained from empty-riser measurements processed identically and was then subtracted directly.

\section{Methodology} \label{Section:Methodology}
\subsection{Near- and Far-Field Beam Expansion Characteristics}

\cref{fig:beam_expansion_fullwidth} presents the irradiance distributions $E(x,y,R)$ obtained by raster-scanning the reference target over a $25\times25$ spatial grid with a pixel size of $\SI{5}{\milli\meter}\times\SI{5}{\milli\meter}$ in ten measurements spanning axial distances between \SI{0.5}{\meter} and \SI{3.9}{\meter}. This included a repeated measurement at $R=\SI{3.6}{\meter}$ with a different raster-plane orientation to assess sensitivity to scan alignment. Background scans acquired over the same spatial domain in the absence of the metallic sphere were subtracted from the target scans. The observed asymmetry between the $x$ and $y$ axes indicates an elliptical beam shape, likely caused by aperture truncation or astigmatic aberrations introduced by the optical components. No secondary maxima were observed within the measured range, consistent with a dominant main-lobe structure. As shown in \cref{fig:beam_expansion}, the extracted half-power beam radii $w_\xi(R)$, $\xi\in\{x,y\}$, increase approximately linearly with range, consistent with paraxial Gaussian beam propagation, i.e.,
\begin{equation}
    w_\xi(R) = w_{0,\xi} + (R-R_0)\tan\psi_\xi,
\end{equation}
where $R_0$ denotes the collimator plane, $w_{0,\xi}=w_\xi(R_0)$ is the fitted beam radius at that plane, and $\psi_\xi$ is the half-divergence angle along principal axis $\xi\in\{x,y\}$. Linear fits to the experimental data yield $\psi_x=0.107^\circ$ and $\psi_y=0.193^\circ$, both satisfying the paraxial (narrow-beam) condition $\psi_\xi\ll1$ \cite{hecht1998optics}.

The scanned distances cover the same interval as subsequent measurements in the riser, inherently compensating for the equivalent near-field effects including beam divergence, diffraction, and system alignment errors. 

A conservative sufficient condition for antenna far-field (Fraunhofer) behavior is given by the Rayleigh distance, $d_{\mathrm{F}}\approx2D_{\mathrm{a}}^{2}/\lambda$, where $D_{\mathrm{a}}$ is the largest transverse dimension of the illuminated footprint on the collimator \cite{AntennaTheoryAnalysisAndDesign2016,NearFieldReflectivityandAntennaBoresightGainCorrectionsforMMWaveRadars}. Using the linear regressions of the half-power footprint radii (semi-axes) $w_x(R)$ and $w_y(R)$ from \cref{fig:beam_expansion}, evaluated at the collimator plane ($R=R_0$), we define $D_{\mathrm{a}}=2\max\{w_{0,x},w_{0,y}\}$. This footprint underfills the mirror (approximately $8\%$ area coverage), giving $d_{\mathrm{F}}\approx\SI{1.8}{\meter}$ at the center frequency. This estimate is conservative because (i) the Rayleigh distance provides a sufficient, rather than necessary, far-field criterion and (ii) the major-axis dimension of an elliptical footprint is used. Consequently, the measurement interval $R=\SIrange{1}{4}{\meter}$ is expected to satisfy $R\gtrsim \SI{1.8}{\meter}$ over most of the range; the smallest ranges may retain residual wavefront curvature. The actual corresponding range dependence of the beam geometry and system response is characterized experimentally, while the spatial variation of the quasi-collimated field occurs over scales much larger than the particle dimensions, supporting a local plane-wave approximation over individual particles.

\subsection{Absolute Power Calibration by Method of Substitution}\label{sec:method_calibration}
\begin{figure*}[!t]
  \centering
  \subfloat[]{%
    \includegraphics[width=1\textwidth]{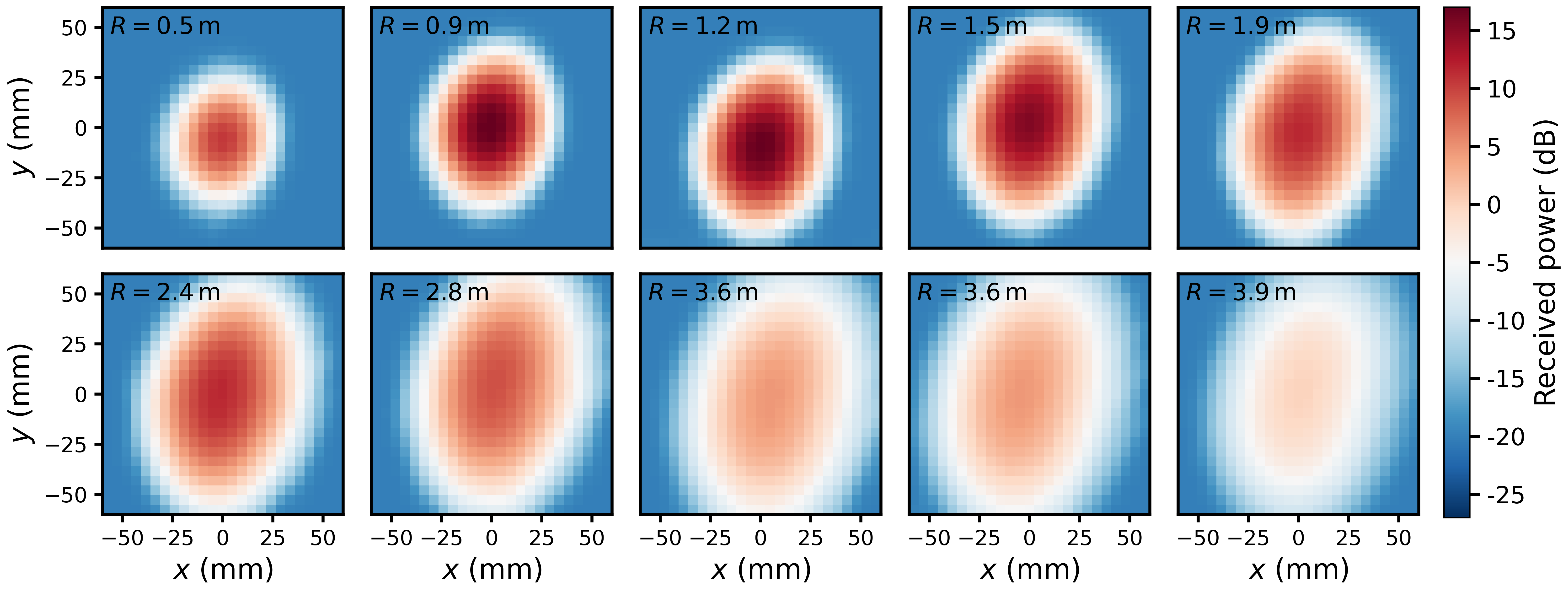}%
    \label{fig:beam_expansion_fullwidth}%
  }\\[1ex]
  \subfloat[]{%
    \includegraphics[width=3.25in]{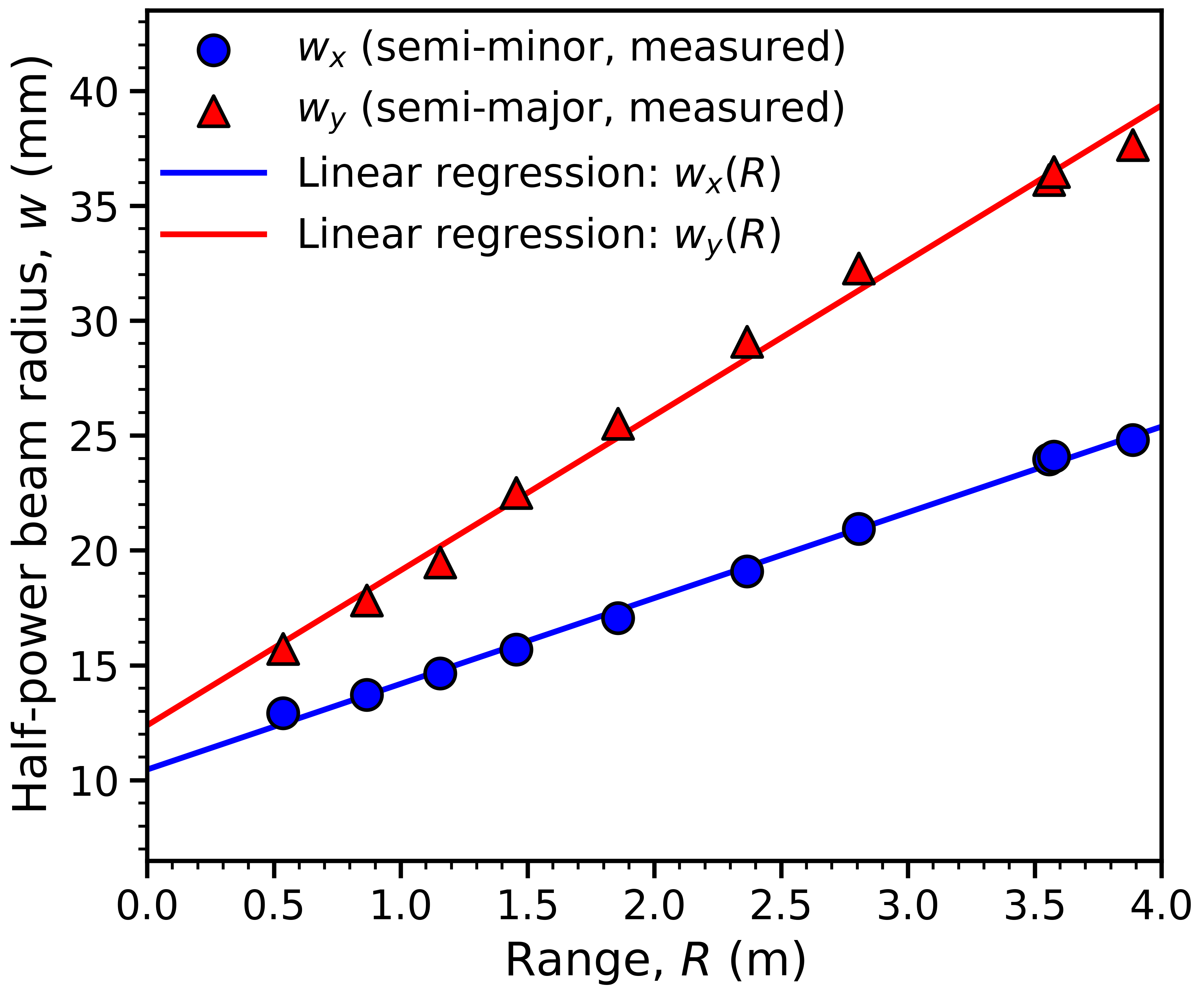}%
    \label{fig:beam_expansion}%
  }
  \hfil
  \subfloat[]{%
    \includegraphics[width=3.25in]{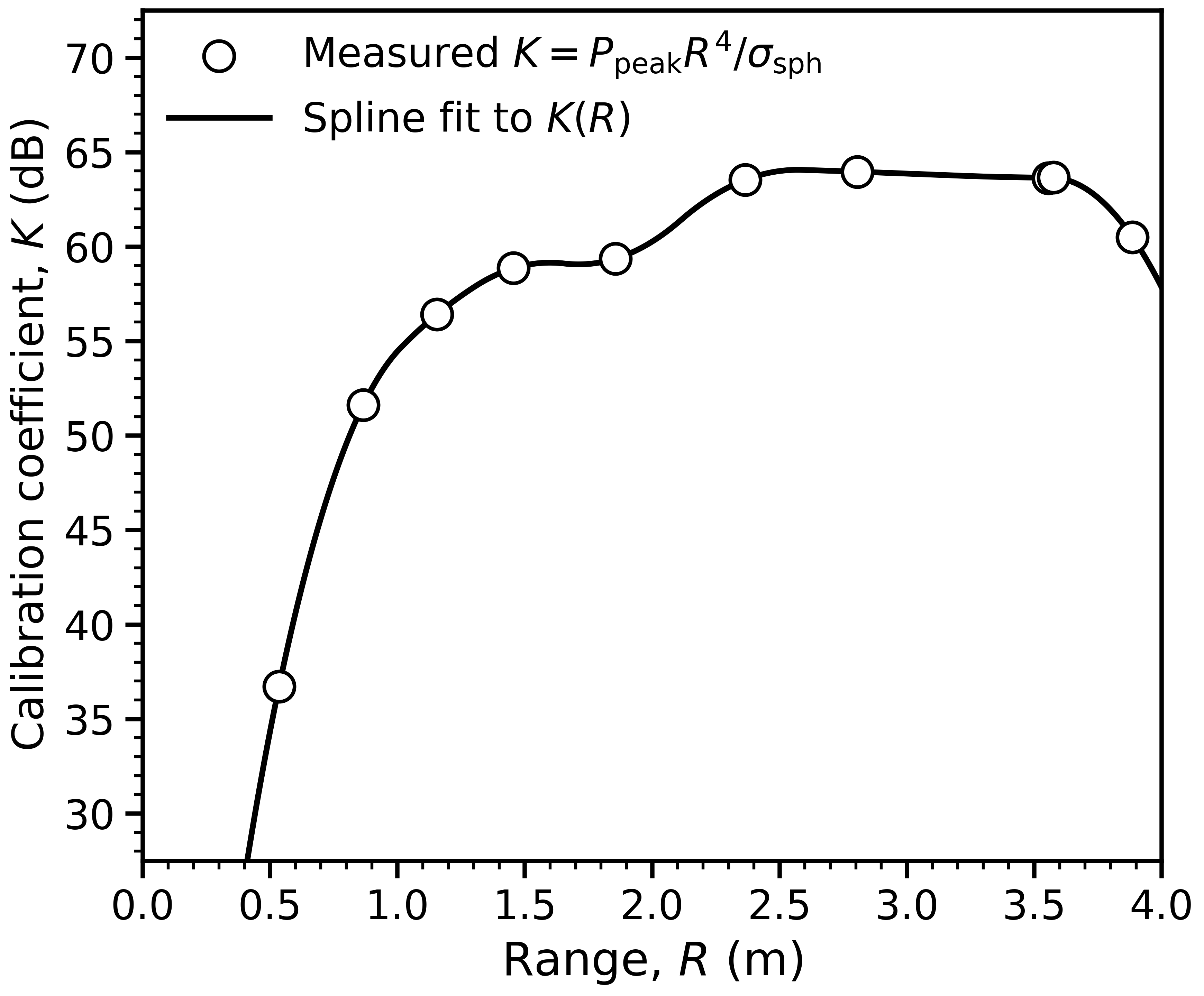}%
    \label{fig:calibration_fit}%
  }
\caption{
(a) Measured two-dimensional beam pattern showing the lateral irradiance distribution $E(x, y)$ at varying propagation distances $R$. (b) Beam expansion of the half-power radii (3-dB semi-axes) $w_x$ and $w_y$. The two measurements at $R=\SI{3.6}{\meter}$ correspond to different raster-plane orientations. Linear regression lines are shown for each axis, with adjusted coefficients of determination of 0.97 and 0.98, respectively. (c) Range-dependent calibration coefficient $K(R) = P_{\mathrm{peak}} R^4 / \sigma_{\mathrm{sph}}$ on a dB scale, where a spline function is applied to match the radar system's effective range resolution.
}
\label{fig:combined_beam_and_calibration}
\end{figure*}

For each measured irradiance distribution (\cref{fig:beam_expansion_fullwidth}), the received peak power $P_{\mathrm{peak}}(R)$ was extracted to compute the calibration coefficient $K(R)=P_{\mathrm{peak}}(R)R^{4}/\sigma_{\mathrm{sph}}$ and quantify its range dependence. A smoothing spline \cite{spline_DIERCKX1975165} was fitted to the measured values of $K(R)$ and evaluated on the radar range grid while suppressing high-frequency noise. The results are shown in \cref{fig:calibration_fit}, where $K(R)$ exhibits three regimes.

Over \SIrange{1}{2.5}{\meter}, $K(R)$ exhibits significant variation due to near-field beam-formation effects. The response stabilizes at approximately \SI{64}{\deci\bel} on the plotted scale over \SIrange{2.5}{3.84}{\meter}, consistent with far-field-like propagation (or reduced near-field effects) over this interval. Beyond \SI{3.84}{\meter}, $K(R)$ decreases monotonically because of bandpass attenuation of the beat frequencies associated with larger ranges. The calibration accounts for this bandpass response. The upper extent of the range-independent response is set by the waveform and filter configuration: the plateau in $K(R)$ over \SIrange{2.5}{3.84}{\meter} corresponds to the range-independent system constant expected once the target is in the far field of the collimated aperture. Atmospheric attenuation over the measurement range is expected to be small relative to scattering-induced extinction from the metallic particle suspension under the present conditions.

\subsection{Stable Inversion of Strongly Attenuated Radar Signals in Particulate Media}
To retrieve the particle number density $n(R)$ from the measured received power $P_r(R)$, we define the lumped quantity $\Phi(R)$, termed the range-calibrated signal, to consolidate all a priori known range-dependent factors:
\begin{equation}
\Phi(R) = \ln\!\left(\frac{P_r(R)\,R^4}{K(R)\,\Delta V(R)}\right).
\label{eq:calibrated_signal}
\end{equation}

Substituting \cref{eq:calibrated_signal} into the radar equation (\cref{SimplifiedRadarEquation}) and differentiating with respect to range yields:
\begin{equation}
\frac{\mathrm{d}\Phi}{\mathrm{d}R} = \frac{1}{\eta(R)}\,\frac{\mathrm{d}\eta}{\mathrm{d}R} - 2\,k(R).
\label{eq:diff_log_signal}
\end{equation}
This expression separates the range dependence of $\Phi(R)$ into a term proportional to the spatial gradient of $\eta(R)$ and a term representing two-way attenuation. The volumetric coefficients $\eta(R)$ and $k(R)$ depend linearly on $n(R)$ according to \cref{eq:scattering_eta,eq:scattering_k}, and substitution into \cref{eq:diff_log_signal} yields a first-order Bernoulli equation in $n(R)$:
\begin{equation}
\frac{\mathrm{d}n}{\mathrm{d}R} - \left(\frac{\mathrm{d}\Phi}{\mathrm{d}R}\right)n = 2\,\sigma_e\, n^2.
\label{eq:riccati_eq}
\end{equation}
Introducing the transformation $u(R)=1/n(R)$ converts \cref{eq:riccati_eq} into a linear first-order equation in $u(R)$, which can be solved by an integrating factor. Imposing a boundary condition at a reference range $R_0$, where $n(R_0)=n_0$ and $\Phi(R_0)=\Phi_0$, yields the forward-propagating solution for $R>R_0$:
\begin{equation}
n(R) = \frac{\exp\bigl(\Phi(R) - \Phi_0\bigr)}{\,n_0^{-1} - 2\,\sigma_e \displaystyle\int_{R_0}^{R} \exp\bigl(\Phi(r) - \Phi_0\bigr)\, \mathrm{d}r}.
\label{eq:forward_solution}
\end{equation}
In strongly attenuating media, \cref{eq:forward_solution} becomes ill-conditioned because the integral term increases with $R$ and can become comparable to $n_0^{-1}$, producing near-cancellation in the denominator. This can induce singular behavior and cause strong sensitivity to perturbations in $\Phi(R)$ and in the boundary value $n_0$. 

For stability, the solution is instead evaluated by backward integration from a far-range boundary. Let $R_f$ denote a far-range reference location where $n(R_f)=n_f$ and $\Phi(R_f)=\Phi_f$. For $R<R_f$, the backward-propagating solution is:
\begin{equation}
n(R) = \frac{\exp\bigl(\Phi(R) - \Phi_f\bigr)}{\,n_f^{-1} + 2\,\sigma_e \displaystyle\int_{R}^{R_f} \exp\bigl(\Phi(r) - \Phi_f\bigr)\, \mathrm{d}r}.
\label{eq:backward_solution}
\end{equation}
In \cref{eq:backward_solution}, the denominator is the sum of non-negative terms and therefore increases monotonically as $R$ decreases from $R_f$ toward the antenna. Consequently, the influence of the far-range boundary value $n_f$ diminishes as the integral term dominates, and the backward formulation mitigates the instability of forward inversion under strong two-way attenuation, enabling stable estimation of range-resolved solids volume fraction profiles from path-attenuated returns within the stated modeling assumptions.

\subsection{Boundary Condition Estimation}
The backward solution in \cref{eq:backward_solution} requires a boundary value for the far-range number density $n_f=n(R_f)$. Because this boundary lies in a region where the signal is strongly attenuated, the same physical assumptions used to derive the inversion are used to estimate $n_f$. The boundary location $R_f$ is selected in the far-range region. Over a short boundary interval $[R_{f\!-\!N},R_f]$ that consists of only a few range bins near $R_f$, the calibrated signal is assumed to be dominated by two-way extinction rather than by spatial variability in the backscattering coefficient, i.e.,
\begin{equation}
\left|\frac{1}{\eta(R)}\frac{\mathrm{d}\eta}{\mathrm{d}R}\right|\ll 2\,k(R),\qquad R\in [R_{f\!-\!N},R_f].
\label{eq:extinction_dominated}
\end{equation}
To assess this approximation, we define the dimensionless ratio $\chi(R)\equiv \left|\eta^{-1}(R)\,\mathrm{d}\eta/\mathrm{d}R\right|/(2\,k(R))$ and evaluate it a posteriori over the boundary bins. Under \cref{eq:extinction_dominated}, \cref{eq:diff_log_signal} implies $\mathrm{d}\Phi/\mathrm{d}R\approx -2\,k(R)$ near $R_f$. Evaluating at $R=R_f$ and using $k(R_f)=\sigma_e n_f$ gives a slope-based estimate:
\begin{equation}
n_f\approx -\frac{1}{2\sigma_e}\left.\frac{\mathrm{d}\Phi}{\mathrm{d}R}\right|_{R=R_f}.
\label{eq:nf_slope_improved}
\end{equation}
Here, the derivative $\mathrm{d}\Phi/\mathrm{d}R$ is estimated from the variation of $\Phi(R)$ across the $N$ boundary bins.

A complementary estimator can be derived from \cref{eq:backward_solution} by introducing a short-interval closure for the attenuation in the exponential weight. Over the boundary interval $[R_{f\!-\!N},R_f]$, we represent the net extinction by the effective boundary value $k(R_f)=\sigma_e n_f$ and evaluate \cref{eq:backward_solution} at $R=R_{f\!-\!N}$, which yields an integral-based estimator:
\begin{equation}
n_f \approx \frac{\exp\bigl(\Phi(R_{f\!-\!N})-\Phi_f\bigr) - 1}{2\,\sigma_e\,\displaystyle\int_{R_{f\!-\!N}}^{R_f}\exp\bigl(\Phi(r)-\Phi_f\bigr)\,\mathrm{d}r}.
\label{eq:nf_integral_improved}
\end{equation}
This method avoids differentiation of $\Phi(R)$ and is therefore less sensitive to random fluctuations in the measured signal. Here, we set $N=2$ (corresponding to $[R_{f\!-\!2},R_f]$) for the boundary interval, and the integral-based estimator \cref{eq:nf_integral_improved} is employed, while \cref{eq:nf_slope_improved} is used as a consistency check.

\section{Measurement Results and Discussion} \label{Section:Results}
\begin{figure*}[!t]
  \centering
  \includegraphics[width=\textwidth]{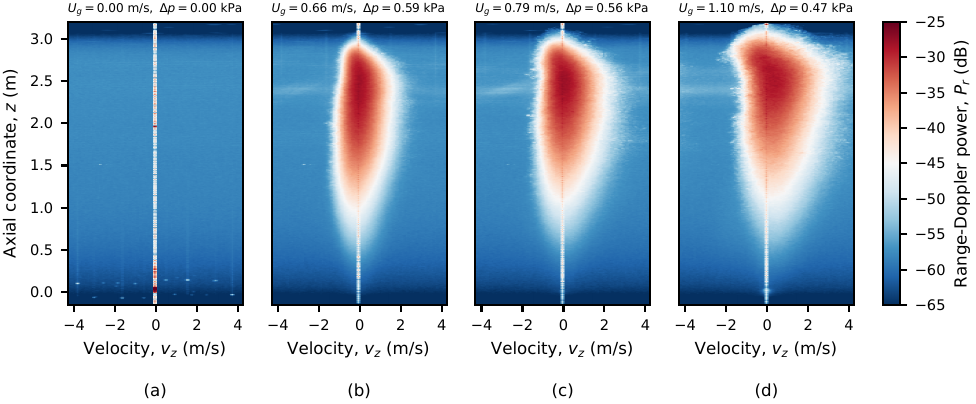}
  \caption{Range--Doppler maps of received power for (a) an empty riser at $U_g=\SI{0}{\meter\per\second}$ and $\Delta p=\SI{0}{\kilo\pascal}$, and for solids-circulation cases at superficial gas velocities of (b) $U_g=\SI{0.66}{\meter\per\second}$, (c) $U_g=\SI{0.79}{\meter\per\second}$, and (d) $U_g=\SI{1.10}{\meter\per\second}$. The corresponding total pressure drops indicate the total cross-sectional mean solids holdup. The distributor-plate surface maps to approximately $z=\SI{0.05}{\meter}$, with the associated double reflection extending to $z\approx\SI{0.15}{\meter}$, and the outlet is centered at approximately $z=\SI{2.75}{\meter}$.}
  \label{fig:DopplerMaps}
\end{figure*}

\begin{figure*}[!t]
  \centering
  \subfloat[]{%
    \includegraphics[width=3in]{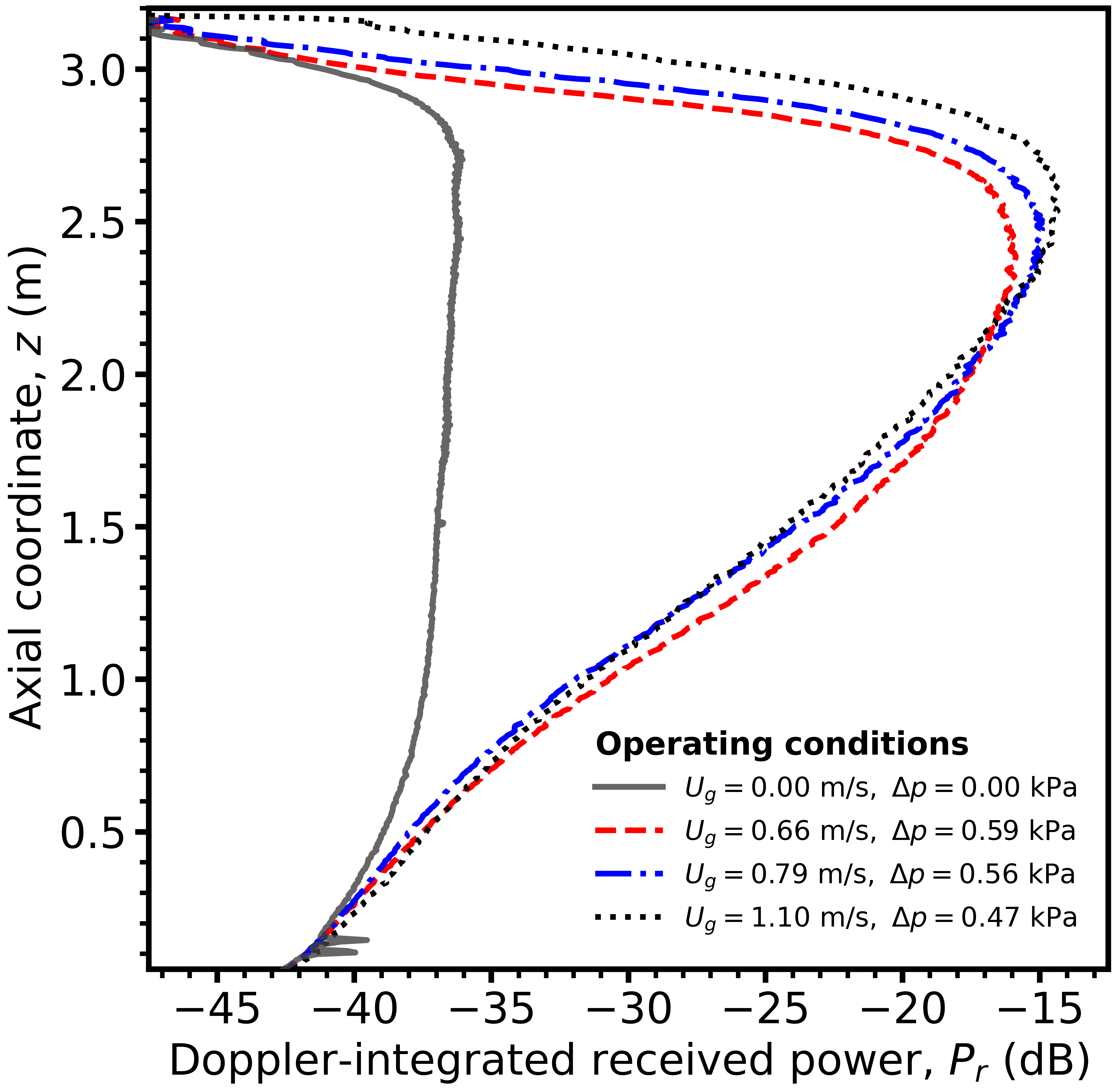}%
    \label{fig:PowerSums}%
  }
  \hfil
  \subfloat[]{%
    \includegraphics[width=3in]{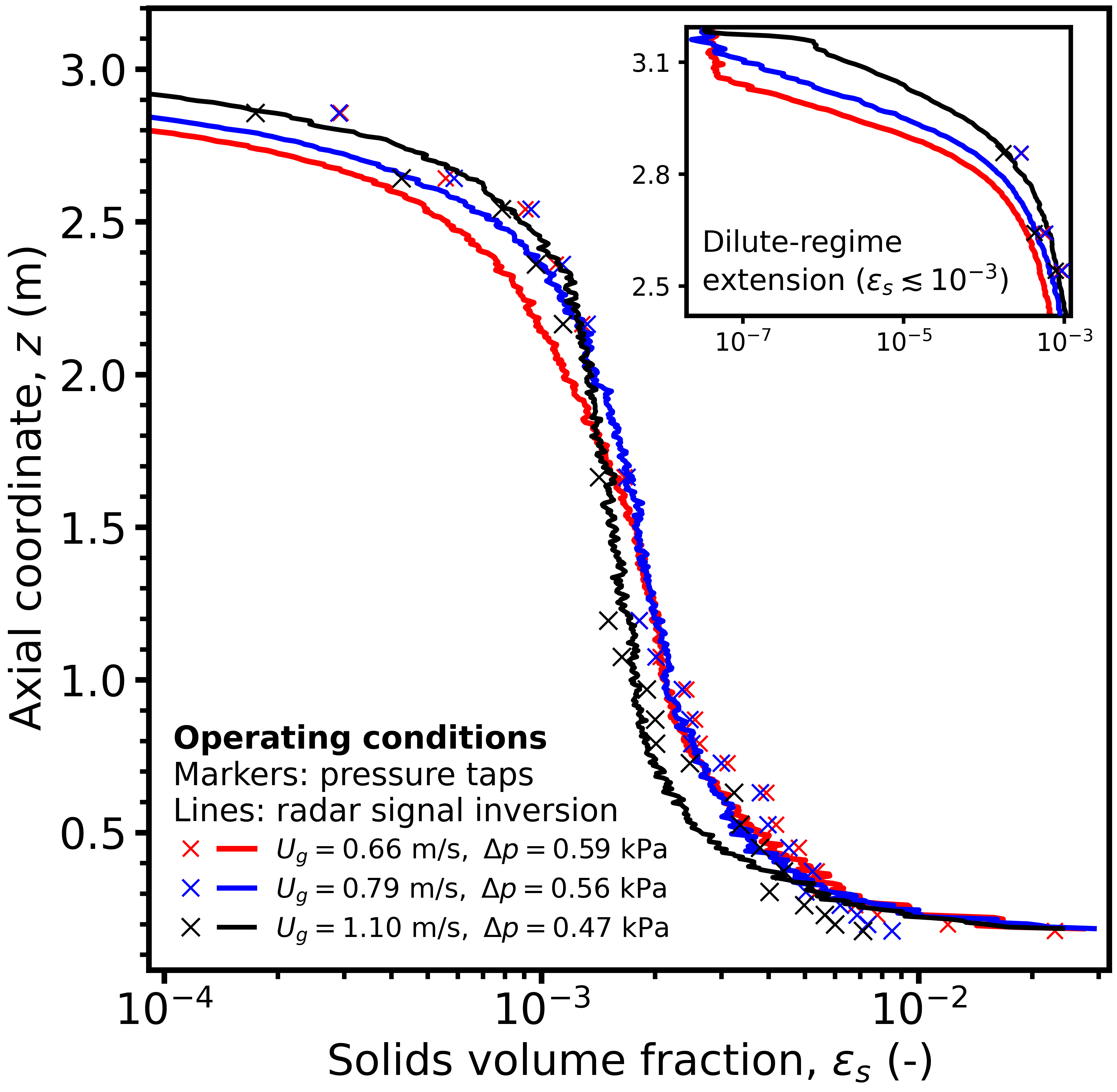}%
    \label{fig:SolidsVolumeFractionProfiles}%
  }
\caption{Axial riser profiles of (a) Doppler-integrated received power $P_r$ and (b) solids volume fraction $\varepsilon_s$, shown for three operating conditions and, in~(a), the empty-riser reference ($U_g=\SI{0}{\meter\per\second}$, $\Delta p=\SI{0}{\kilo\pascal}$). In~(a), the increase in $P_r$ relative to the empty-riser baseline reflects particle backscatter along the radar measurement axis. In~(b), solid curves denote the local solids volume fraction profiles retrieved by radar-signal inversion; cross markers denote the corresponding estimates derived from differential-pressure measurements. The inset in~(b) magnifies the dilute region ($\varepsilon_s \lesssim 10^{-3}$) extending from the outlet elevation to the riser roof, where pressure data are not available.}
  \label{fig:profiles_twopanels}
\end{figure*}

\cref{fig:DopplerMaps,fig:profiles_twopanels} present the measured range--Doppler maps, the Doppler-integrated power profiles, and the solids volume fraction profiles retrieved by the inversion, together with the pressure-based estimates used for comparison.
\cref{fig:DopplerMaps} shows the time-averaged range--Doppler maps acquired for the empty-riser reference and the three operating conditions. The horizontal axis gives the axial solids velocity $v_z$, the vertical axis gives the riser-relative axial height $z$, and the color scale represents the received power (in dB) at each height and velocity. 

\cref{fig:DopplerMaps}\textcolor{blue}{a} shows the empty-riser reference, measured without inlet gas flow or particles, which is used for background subtraction of non-particulate returns before the radar-signal inversion. \cref{fig:DopplerMaps}\textcolor{blue}{b--d} show the three operating cases with suspended copper powder, with $U_g$ increasing from left to right and the corresponding pressure drop indicating the total solids holdup in the riser.

Profiles are reported in the riser-relative axial coordinate presented in \cref{Riser_reference_system}. The distributor-plate surface in \cref{fig:DopplerMaps}(a) maps to approximately $z=\SI{0.05}{\meter}$, with the associated double-reflection feature extending to $z\approx\SI{0.15}{\meter}$, while the HDPE roof maps to $z\approx\SI{3.15}{\meter}$. In the near-bottom region ($z\leq\SI{0.15}{\meter}$ in \cref{fig:DopplerMaps}\textcolor{blue}{a}), the range bins are dominated by near-zero-Doppler specular returns, including leakage into adjacent Doppler bins, and are therefore excluded from the quantitative analysis.

As shown in \cref{fig:DopplerMaps}\textcolor{blue}{b--d}, increasing gas velocity is accompanied by a lower total pressure drop $\Delta p$, indicating reduced total solids holdup in the riser, and by broader Doppler spectra in the radar return. The spectral broadening is most pronounced for upward solids motion ($v_z>0$), whereas the downward component ($v_z<0$) is less affected. The Doppler-resolved return associated with upward-moving solids indicates that upward transport is the dominant direction of motion. However, the weaker downward-moving component indicates local counter-current solids motion, consistent with solids back-mixing within the core region \cite{ToveHenrik_DNS, RadarBased_Wu_2024}.

The outlet centerline is located at approximately $z=\SI{2.75}{\meter}$. Near and above this elevation, stronger returns from downward-moving solids emerge in the range--Doppler maps, consistent with flow redirection imposed by the outlet geometry. Below the outlet centerline, the return power remains dominated by upward-moving solids. At higher gas velocities, detectable solids returns extend farther into the region above the outlet centerline, indicating increased solids transport into the upper outlet-affected region. These observations are consistent with outlet-induced solids redirection reported for CFB risers \cite{WU2023_SolidsSeparationEfficiencyCFB}; however, a detailed fluid-dynamic analysis of the outlet region is outside the scope of this work.

Toward the top of the maps $z\gtrsim\SI{3.0}{\meter}$ ($R\lesssim\SI{1.15}{\meter}$), the received power decreases gradually. This follows from the soft roll-off of the digital high-pass filter applied to the beat spectrum (\cref{RadarSystemDescription}) and is instrumental in origin rather than a feature of the riser flow. Because the roll-off onset lies just below the riser roof, the topmost dilute returns fall within it and are interpreted accordingly below.

\cref{fig:profiles_twopanels} shows the range-resolved Doppler-integrated return-power profiles and the corresponding solids volume fraction profiles obtained from the radar-signal inversion, together with pressure-based estimates. \cref{fig:PowerSums} shows the total backscattered power profiles $P_r(z)$, formed by subtracting the empty-riser reference in \cref{fig:DopplerMaps}\textcolor{blue}{a} from the operating Doppler maps in \cref{fig:DopplerMaps}\textcolor{blue}{b--d} and integrating the remaining return over Doppler velocity at each range. These profiles isolate the particle-associated backscattered power and are used as input to the inversion, yielding the radar-based solids volume fraction profiles in \cref{fig:SolidsVolumeFractionProfiles}.

From \cref{fig:PowerSums}, the Doppler-integrated received power exhibits the strongest height dependence in the lower riser, whereas the highest power levels occur in the upper section. Above the distributor-plate double reflections ($z\geq\SI{0.15}{\meter}$), the particle-associated return remains above the empty-riser background up to the riser roof, corresponding to a signal-to-background ratio greater than unity. Below the outlet region ($z\leq\SI{2.5}{\meter}$), the maximum power-decay rate is approximately $\SI{15}{\deci\bel\per\meter}$ in all cases.

\cref{fig:SolidsVolumeFractionProfiles} shows the solids volume fraction profiles retrieved by inversion with the far-range boundary set to $R_f=\SI{3.93}{\meter}$, corresponding to $z_f=\SI{0.185}{\meter}$. This boundary location satisfies $z_f\geq\SI{0.15}{\meter}$ and thereby excludes residual static-clutter leakage from the distributor-plate region in the empty-riser background. The retrieved $\varepsilon_s$ decreases monotonically with height, from values of order $10^{-2}$ in the lower riser to values of order $10^{-3}$ through the mid-riser, followed by a further decrease toward the riser roof.

Sensitivity tests varying the far-range boundary over $R_f=\SIrange{3.65}{3.95}{\meter}$ (corresponding to $z_f\approx\SIrange{0.17}{0.47}{\meter}$) and the boundary-interval width over $N=2$--$10$ range bins produced only minor changes in the retrieved profiles and in the estimated $n_f$ for both the slope- and integral-based estimators. A posteriori, $\chi(R)=\mathcal{O}(10^{-1})$ over $[R_{f-N},R_f]$ supports \cref{eq:extinction_dominated} and $\mathrm{d}\Phi/\mathrm{d}R\approx-2\,k(R)$ in the boundary bins; agreement between the estimates in \cref{eq:nf_slope_improved,eq:nf_integral_improved} provides an internal consistency check on the boundary selection. The far-range boundary $R_f=\SI{3.93}{\meter}$ lies beyond the conservative far-field distance $d_{\mathrm{F}}\approx\SI{1.8}{\meter}$, such that the backward solution is initialized where the far-field criterion is satisfied. At $R\lesssim d_{\mathrm{F}}$ ($z\gtrsim\SI{2.3}{\meter}$), this criterion is not satisfied; although the range-dependent beam geometry and system response are characterized experimentally through $\mathcal{A}(R)$ and $K(R)$, residual near-field effects may increase the retrieval uncertainty. Because the inversion propagates toward decreasing $R$, such effects enter only after the solution reaches this interval and do not propagate into retrievals at larger ranges.

The far-range inversion boundary ($R_f=\SI{3.93}{\meter}$) and part of the sensitivity interval lie beyond the bandpass-filter range $R_{\mathrm{BPF}}=\SI{3.84}{\meter}$ ($z=\SI{0.275}{\meter}$), where the upper-frequency roll-off of the physical bandpass filter attenuates the densest, lowest part of the riser ($z\lesssim\SI{0.275}{\meter}$). This roll-off reduces the measured Doppler-map intensity there but does not by itself invalidate the retrieved magnitudes: the external-substitution calibration determines the range-dependent system response $K(R)$, which includes the bandpass response, so the inversion magnitude is constrained jointly by the far-range boundary, the Mie scattering model, and the calibrated response rather than by the raw map intensity. The densest retrievals ($z\lesssim\SI{0.275}{\meter}$) are therefore interpreted with caution, as the reduced signal-to-noise ratio increases their uncertainty, while remaining quantitatively meaningful within the calibrated framework.

The radar-retrieved $\varepsilon_s$ profiles are consistent with the pressure-derived estimates across the three operating conditions. The concentration increases toward the base while the received power decreases there (\cref{fig:PowerSums}): the dense lower suspension imposes strong two-way attenuation that the backward inversion compensates, recovering the higher local concentration from the reduced return. Toward the upper riser, cumulative particulate attenuation is lower because of the shorter propagation path. The inset of \cref{fig:SolidsVolumeFractionProfiles} shows that the radar retains sensitivity to dilute concentrations of order $10^{-6}$, where pressure-derived estimates become unreliable. Toward the riser roof, the lowest retrieved values fall below $10^{-7}$, reaching approximately $5\times10^{-8}$. Although this region coincides with the soft roll-off of the digital high-pass filter ($z\gtrsim\SI{3.00}{\meter}$, $R\lesssim\SI{1.15}{\meter}$), its range-dependent response is included in the experimentally determined $K(R)$. These lowest values are reported as observed retrievals rather than as a quantitative detection floor because no independent reference or detection-limit criterion is available in this concentration interval.

The independent-scattering closure of \cref{sec:ensemble_averaging} is assessed a posteriori from the retrieved profiles. Across the measured size distribution, the size parameter remains below unity ($x(D_{90})\approx0.24$ at $f_c$), so all particles are electrically small and the Mie cross-sections in \cref{eq:scattering_eta,eq:scattering_k} lie in the dipole-dominated regime; the ensemble asymmetry parameter $g\approx-0.38$ is consistent with the backscatter-enhanced response expected of electrically small, highly conducting spheres \cite{AsymmetryParameterConductingSphere2019,AbsScatBySmallParticles_CH5}. The low solids volume fraction ($\varepsilon_s\lesssim2\cdot10^{-3}$ for $z\gtrsim\SI{0.4}{\meter}$, \cref{fig:SolidsVolumeFractionProfiles}) keeps the suspension dilute over this interval and, together with the exponential speckle statistics of \cref{fig:fmcw_sim}\textcolor{blue}{b}, supports the incoherent power addition of \cref{IncoherentScatteringEquation}. Dependent (near-field) coupling between scatterers is a separate limitation: evaluating \cref{eq:mean_distance} across the profiles, the mean inter-particle separation relative to the wavelength ranges from $\langle d\rangle/\lambda\approx5\times10^{-2}$ in the densest range bins to $\sim\!2.5$ near the riser roof. Dependent-scattering effects therefore cannot be excluded where $\langle d\rangle\lesssim\lambda$; this limitation is most pronounced in the denser lower riser and relaxes progressively toward the dilute upper region, and it is not represented by the independent-particle closure.

The range-resolved return is modeled as first-order backscatter (single scattering): radiation reaching the receiver after more than one scattering event is neglected, whereas extinction of the directly propagating field is retained through the two-way attenuation factor. The two-way path-integrated extinction exceeds unity below $z\approx\SI{2.3}{\meter}$, which reflects the strong path attenuation that the inversion explicitly models and does not by itself imply a large multiply-scattered return. For electrically small particles, only wide-angle multiple scattering is possible, and for a finite beam its detected contribution is estimated by the ratio of the footprint radius to the extinction mean free path, $X_r=\sqrt{w_x w_y}/(1/k)$~\cite{Battaglia2010MSReview,Kobayashi2005FiniteBeam}. Over the relevant measurement interval ($\SI{0.275}{\meter}\lesssim z\lesssim\SI{3.0}{\meter}$), $X_r\leq0.03\ll1$, so laterally scattered radiation largely leaves the narrow field of view before re-scattering toward the receiver. The computed phase function shows no dominant forward lobe ($g\approx-0.38$; scattering is in fact backward-enhanced, as expected for electrically small, highly conducting spheres), so the small-angle forward multiple scattering that can accumulate along the propagation path in media of large particles does not arise here. Consequently, for the present narrow-beam monostatic geometry, higher-order returns are expected to be suppressed.

The agreement between the radar- and pressure-derived $\varepsilon_s$ across the three operating conditions---including the dense lower riser (\cref{fig:SolidsVolumeFractionProfiles}), where the path-integrated extinction is largest and the inter-particle spacing smallest---provides a posteriori support for the adequacy of the adopted scattering closure under the investigated conditions, and indicates that the neglected dependent- and higher-order scattering contributions do not dominate the retrieval error. It cannot, however, isolate or quantify the individual electromagnetic effects. Additional uncertainty arises from the assumptions of range-invariant size statistics and a spatially homogeneous particle field, which do not capture height-dependent variation in the particle-size distribution or particle clustering within the scattering volumes. Beyond the present conditions, both the external calibration and the retrieved extinction apply to a dry, ambient-temperature gas phase, for which molecular absorption within the swept band is expected to be small relative to the particulate extinction. Because the inversion attributes the full path-integrated extinction to the particulate phase, any non-negligible gas-phase absorption, as encountered in humid or high-temperature process gas, would instead be interpreted as particulate extinction and would have to be quantified independently.

\section{Conclusions}
\label{Section:Conclusions}
This work developed and experimentally validated an inversion framework for retrieving range-resolved solids concentration from monostatic \SI{0.34}{\tera\hertz} FMCW pulse--Doppler radar measurements under strong cumulative attenuation. The path-integrated radar equation reduces to a Bernoulli differential equation with a closed-form solution, which is evaluated by backward integration from a far-range boundary to avoid the ill-conditioning of forward integration under strong path-integrated extinction. Local concentrations are thereby recovered from path-attenuated returns. The method requires no case-specific parameter fitting to reference concentration measurements; instead, the model is closed using ensemble-averaged Mie cross-sections evaluated over the measured polydisperse particle-size distribution and an absolute range-dependent system response determined empirically through external substitution with an electrically large metallic sphere.

Measurements in a \SI{3.1}{\meter}-tall CFB riser containing copper powder and air at ambient conditions produced radar-derived profiles consistent with differential-pressure-derived estimates across three distinct operating conditions. The radar remained sensitive to solids volume fractions of order $10^{-6}$ in the dilute upper riser, where pressure-based estimates are unreliable. The agreement between pressure and radar estimates suggests that dependent- and higher-order-scattering effects do not dominate the retrieval error under the tested conditions, although the comparison does not isolate or quantify these effects.

Because the framework does not fit model parameters to pressure-derived concentrations, it supports application to other viewing geometries, including lateral measurements across the riser, for which pressure measurements cannot provide equivalent spatial information. Such transfer, however, requires verification of the assumptions underlying the boundary condition and, for high-temperature applications, explicit treatment of gas-phase absorption as a separate contribution to path attenuation. Application to other particulate materials additionally requires the corresponding scattering cross-sections and particle-size distributions. Establishing performance limits for industrial units requires further evaluation at greater propagation lengths, higher solids holdup, and stronger gas-phase attenuation. Because line-of-sight access is required only through a single small aperture, stable inversion under stronger attenuation may support nonintrusive monitoring where intrusive or optical access is impractical.

\section*{Acknowledgment} 
This work was funded by the Swedish Research Council under the project ``Disclosing the particle scale to enable reliable full-scale simulations of gas-solids flow'' (2023-03970).

\bibliographystyle{IEEEtran}
\bibliography{refs}

\vfill

\end{document}